\documentclass[aps,floats,pre,showpacs
]{revtex4}

\newcommand{\Z} {{\cal Z}}
\newcommand{\tL} {{\cal L}}
\newcommand{\mF}{\mathscr{F}}
\newcommand{\mL}{\mathscr{L}}

\usepackage{amsfonts,amsmath,mathrsfs} \usepackage{bm} \usepackage{dcolumn}
\usepackage{epsfig} \usepackage{latexsym}

\begin{document}

\title{Electron fractionalization and unconventional order parameters of the $t$-$J$ model}

\author{Peng Ye$^{1}$, Chushun Tian$^{1,2}$, Xiaoliang Qi$^3$ and Zhengyu Weng$^1$}

\affiliation{$^{1}$ Institute for Advanced Study, Tsinghua
University,
Beijing, 100084, P. R. China\\
$^{2}$ Institut f\"{u}r Theoretische Physik, Universit\"{a}t zu K\"{o}ln, D-50937 K\"{o}ln, Germany \\
$^{3}$ Department of Physics, Stanford University, Stanford, CA
94305, USA}



\begin{abstract}
{\rm In the $t$-$J$ model, the electron fractionalization is unique
due to the non-perturbative phase string effect. We formulated a
lattice field theory taking this effect into full account. Basing on
this field theory, we introduced a pair of Wilson loops which
constitute a complete set of order parameters determining the phase
diagram in the underdoped regime. We also established a general
composition rule for electric transport expressing the electric
conductivity in terms of the spinon and the holon conductivities.
The general theory is applied to studies of the quantum phase
diagram. We found that the antiferromagnetic and the superconducting
phases are dual: in the former, holons are confined while spinons
are deconfined, and {\it vice versa} in the latter. These two phases
are separated by a novel phase, the so-called Bose-insulating phase,
where both holons and spinons are deconfined and the system is
electrically insulating.}
\end{abstract}

\maketitle

\section{Introduction}
\label{intro}

One of the canonical notions in the theory of thermal phase
transition is the Landau-Ginzburg-Wilson (LGW) symmetry-breaking
paradigm. For a second order phase transition, one utilizes a few
``order parameters'' to characterize different symmetries of the
phases and thus a critical theory is in terms of fluctuations of
these order parameters, which drive the system towards a critical
point. In the past few years, whether the LGW paradigm may be
extended to quantum phase transitions has become one of the most
fascinating topics in studies of strongly correlated electronic
materials. In fact, there has been substantial investigations (e.g.,
Refs.~\cite{Fisher04,Fisher04a,Balents05,Senthil07}) showing that
the conventional LGW paradigm fails to describe diverse quantum
phase transitions. In particular, a microscopic mechanism that
violates the LGW paradigm is the so-called ``fractionalization'',
discovered, say, in the study of two-dimensional quantum magnets
\cite{Fisher04}. It suggests a very general scenario of quantum
phase transitions that has been currently under intensive
experimental and theoretical investigations. Basically, a quantum
particle may be effectively fractionalized into a few degrees of
freedom interacting via emergent gauge field, and some
fractionalized degrees of freedom appear at the critical point but
{\it not} in the stable phases at each side of the critical point.

Thus, as correlated electronic systems are concerned, a fundamental
issue is: how does the electron fractionalization, where electrons
are effectively fractionalized into the spin and the charge degrees
of freedom at short scales, lead to various phases and the phase
transitions between them? In one dimension, it has been well
established that the spin-charge separation results in the
Tomonaga-Luttinger liquid. In contrast, less has been known about
the interplay between the electron fractionalization and quantum
phase transitions, notably in the two-dimensional doped Mott
insulator \cite{Anderson87} which is widely accepted to be the
prototype of high temperature superconductors
\cite{Lee06,Fisher00,Sachdev03}. Notwithstanding, experimental and
theoretical analysis indicating the electron fractionalization to be
at the root of low-dimensional strongly correlated electronic
systems has been well documented. In fact, there exist two classes
of experiments on cuprate materials, namely the large Nernst effect
in thermoelectric transport measurements \cite{Ong04} and the
periodic modulations of electronic density of states in scanning
tunneling microscopy studies \cite{Takagi04}. It has been pointed
out \cite{Balents05} that a LGW-type theory trying to reconcile
these two phenomena cannot be, at least, extended to zero
temperature. Instead, substantial theoretical analysis shows that
the electron fractionalization necessarily leads to new quantum
phases \cite{Lee06}. This opinion has been reiterated very recently
by Zaanen and Overbosch \cite{Zaanen09}, who pointed out that the
drastic change in the nature of quantum statistics upon doping -- a
direct manifestation of the electron fractionalization -- may be the
key to understanding physics of the pseudogap phase. Central to
these studies are a number of fundamental issues. The most prominent
ones include: {\it What kind of quantum statistics do the emergent
degrees of freedom obey? How do they interact with each other? How
does the electron fractionalization turn an antiferromagnet into a
superconductor upon doping? What does the phase diagram look like?}

While to investigate these issues for general doped Mott insulators
proves to be highly challenging, in this work, we switch to a
limiting case where the on-site Coulomb repulsion is strong enough,
namely the $t$-$J$ model (on a bipartite lattice), and will present
an analytical study of the above problems. Our starting point is an
exact non-perturbative result for this model, the so-called {\it
phase string effect} discovered some time ago \cite{Weng97}.
Specifically, the motion of a hole may leave a trace on the lattice
plane upon colliding with spins. The phase string effect means that
such two nearby propagating paths of the hole, though possessing
(almost) identical amplitudes, may substantially differ in the
phases by $\pi$ (namely in the overall sign). Such a sign structure
accounts for the parity of the number of hole--spin collisions and
is thereby intrinsic to individual paths. Most importantly, with
this effect being taken into full account, it turns out (to be
detailed in this work) that an electron is uniquely, {\it rather
than arbitrarily}, fractionalized into two bosonic constituents
namely the holon and the spinon, satisfying a mutual statistical
interaction. Basing on this exact result, in this work, we will
study--at a full microscopic level--how a phase diagram flows out of
electron fractionalization.

The main results of this work were briefly reported in our earlier
publication \cite{Ye11}. The present paper is devoted to reporting
the technical details and the remaining is organized as follows. For
the self-contained purposes we will briefly review the phase string
effect in Sec.~\ref{sec_phasestring} and, particularly, introduce
the emergent mutual statistical interaction. In
Sec.~\ref{sec_latticeMCS}, we will formulate a general lattice field
theory basing on the phase string effect and introduce a pair of
unconventional order parameters. Using this general theory, we will
present a general result for electric transport in
Sec.~\ref{Sec_linearresponse}. In particular, we will prove a
composition rule that expresses the electric conductivity in terms
of holon and spinon conductivities. Armed with the pair of
unconventional order parameters and the composition rule, we will
explore the quantum phase diagram in Sec.~\ref{sec_phase}.
(Throughout this paper, we shall focus on the underdoped regime.)
Finally, we will conclude in Sec.~\ref{sec_conclusion}. Some of the
technical details are regelated to Appendices~\ref{current}-\ref{proof}.

\section{Phase string effect: a short review}
\label{sec_phasestring}

For the self-contained purpose we shall start from a brief
exposition of the exact phase string effect \cite{Weng07} for the
$t$-$J$ model ($t,J>0$)
\begin{eqnarray}
H =
-t \sum_{\langle i,j\rangle}\sum_{\sigma }\,
\left({c}_{i\sigma}^\dagger {c}_{j\sigma} + {\rm h.c.}\right)+ J
\sum_{\langle i,j\rangle} \left(S_i \cdot {S}_j - \frac{{n}_i
{n}_j}{4}\right)
\equiv H_t + H_J
\label{tJ}
\end{eqnarray}
on a square lattice, where ${\rm h.c.}$ stands for the hermitian
conjugate. In this Section, in particular, we will discuss the
physical picture of the mutual statistical interaction between
holons and spinons, as well as its mathematical origin. In
Eq.~(\ref{tJ}), $c_{i\sigma}^\dagger$ ($c_{i\sigma}$) is the
(fermionic) electron creation (annihilation) operator on site $i$
with the spin polarization $\sigma=\downarrow,\uparrow$\,, $n_i
\equiv \sum_\sigma c_{i\sigma}^\dagger c_{i\sigma}$ is the
occupation number operator, $S_i$ is the spin operator, and $\langle
i,j\rangle$ stands for the link between two nearest neighbors $i$
and $j$. Note that throughout this work, we shall not distinguish
notationally the difference between the operators and the numbers.
The wave function is restricted on the Hilbert space such that no
sites are occupied by more than one electron--the
no-double-occupancy constraint. Eq.~(\ref{tJ}) together with the
no-double-occupancy constraint serve as the exact starting point of
the present work.

\subsection{Bosonization: phase-string representation}
\label{bosonization}

In Eq.~(\ref{tJ}) the first ($H_t$) and the second ($H_J$) term
describe the charge hopping and the spin-flip process, respectively.
The latter is reduced into the Heisenberg model in the undoped
(half-filling) case where ${n}_i = 1$\,. For this case, long time
ago Marshall showed that the matrix element of the superexchange
Hamiltonian must satisfy a sign rule \cite{Marshall55}. That is,
there exists a complete set of spin bases $\{|\phi\rangle\}$ such
that given arbitrary spin configurations $\phi$ and $\phi'$\,,
\begin{equation}
\langle\phi'|H_J|\phi\rangle \leq 0 \,. \label{signrule}
\end{equation}
For doped antiferromagnets the Marshall sign rule (\ref{signrule})
may be trivially realized within the Schwinger-boson
(${b}_{i\sigma}^\dagger,{b}_{i\sigma}$), slave-fermion
(${f}_i^\dagger,{f}_i$) formalism, provided that the spin-hole bases
$\{|\phi\rangle\}$ are chosen appropriately. Under these bases the
superexchange and hopping Hamiltonian are written as \cite{Weng97}
\begin{eqnarray}
H_J &=& -\frac{J}{2} \sum_{\langle i,j\rangle}\sum_{\sigma\sigma'}\,
{b}_{i\sigma}^\dagger {b}_{j-\sigma}^\dagger {b}_{j-\sigma'}
{b}_{i\sigma'} \,, \label{Hj} \\
H_t &=& -t \sum_{\langle i,j\rangle}\sum_\sigma\, \sigma {
f}_i^\dagger {f}_j {b}_{j\sigma}^\dagger {b}_{i\sigma} + {\rm
h.c.}\,. \label{Ht}
\end{eqnarray}
Note that the sign of $H_J$ is now negative. Here the ``holon''
creation operator ${f}_i^\dagger$ and the ``spinon'' annihilation
operator ${b}_{i\sigma}$ commute with each other, and satisfy the
no-double-occupancy constraint:
\begin{equation}
{f}_i^\dagger {f}_i + \sum_\sigma {b}_{i\sigma}^\dagger
{b}_{i\sigma} =1 \,. \label{nodouble}
\end{equation}

The sign structure in $H_t$ (arising from hole-spin exchange)
results in peculiar properties. Technically, it prohibits a
perturbative expansion of the resolvent: $(E-H_J-H_t)^{-1}$ even in
$t/J\ll 1$. Physically, it accounts for the fact
\cite{Weng97,Weng07} that the holon, upon hopping to some site
occupied instantly by a $\uparrow$ ($\downarrow$)-spin, acquires a
sign $+1$ ($-1$). (As such, the Marshall sign rule is incompatible
with the hopping process.) Therefore, two holon paths close to each
other, though having (almost) identical amplitudes, may differ in
the sign, because the parity of the number of
hole--$\downarrow$-spin collisions of each paths may be different.
This is the so-called {\it phase string effect} and possesses
non-perturbative nature.

Before further proceeding, it is necessary to take this
non-perturbative phase-string effect into full account. Such a task
was fulfilled in Ref.~\cite{Weng97} by introducing the so-called
phase string transformation:
\begin{equation}
|\psi\rangle \rightarrow e^{i{\Theta}} \, |\psi\rangle \,, \qquad
{O} \rightarrow e^{i{\Theta}} {O} e^{-i{\Theta}} \,, \label{unitary}
\end{equation}
which exactly ``gauges away'' the phase string arising from the
holon hopping. In Eq.~(\ref{unitary}) the phase string operator
${\Theta}$ is defined as
\begin{equation}
{\Theta}\equiv -\sum_{i,j}\, {n}_i^h\,\theta_{ij}\,
{n}^s_{j\downarrow} \,, \qquad \theta_{ij}\equiv {\rm Im} \ln
(z_i-z_j) \in [-\pi,+\pi) \,, \label{operator}
\end{equation}
where ${n}^h_i$ and ${n}^s_{i\sigma}$ are the holon and the spinon
occupation number operator, respectively. Note that the
two-dimensional lattice spans a complex plane and $z$ is the
position of the spinon (holon) in this complex plane. Moreover, the
no-double-occupancy constraint implies that the summation excludes
automatically the term with $i=j$\,.


Let us now apply the phase-string transformation (\ref{unitary}) to
Eqs.~(\ref{Hj}) and (\ref{Ht}). Introduce the bosonic operator
${h}_i^\dagger\,, {h}_i$\,:
\begin{eqnarray}
{h}_i^\dagger \equiv {f}_i^\dagger \, \exp\left(-i\sum_{l\neq
i}\theta_{il}{n}_l^h\right)\,, \qquad {h}_i
\equiv\exp\left(i\sum_{l\neq i}\theta_{il}{n}_l^h\right) \, {f}_i
\label{hdefinition}
\end{eqnarray}
which preserves the no-double-occupancy constraint:
\begin{equation}
{h}_i^\dagger {h}_i + \sum_\sigma {b}_{i\sigma}^\dagger
{b}_{i\sigma} =1 \,. \label{nodouble1}
\end{equation}
Then, $H_{J,t}$ are transformed to
\begin{eqnarray}
H_J & \rightarrow & -\frac{J}{2} \sum_{\langle
i,j\rangle}\sum_{\sigma\sigma'}\, e^{i\sigma A_{ij}^h}\,
b_{i\sigma}^\dagger b_{j-\sigma}^\dagger e^{i\sigma'
A_{ji}^h} \, b_{j-\sigma'} b_{i\sigma'}\,, \label{Hj1}\\
H_t & \rightarrow & -t \sum_{\langle i,j\rangle}\sum_\sigma\, e^{i
A_{ij}^s} \, h_i^\dagger h_j e^{i\sigma A_{ji}^h}\,
b_{j\sigma}^\dagger b_{i\sigma} + {\rm h. c.}\,, \label{Ht1}
\end{eqnarray}
where the gauge fields $A_{ij}^{s,h}$ given by
\begin{eqnarray}
A_{ij}^s &  \equiv  & \frac{1}{2} \sum_{l\neq i,j}\,
\left(\theta_{il}-\theta_{jl}\right)
\left(n^s_{l\uparrow}-n^s_{l\downarrow}\right) \,\, {\rm mod}\, 2\pi \,,\label{gauge3}\\
A_{ij}^h &  \equiv  & \frac{1}{2} \sum_{l\neq i,j}\,
\left(\theta_{il}-\theta_{jl}\right) n^h_l \,\, {\rm mod}\, 2\pi
\label{gauge4}
\end{eqnarray}
are angular variables. Eqs.~(\ref{Hj1})-(\ref{gauge4}) constitute
the exact bosonization of the original $t$-$J$ model (\ref{tJ})
subject to the no-double-occupancy constraint.

\subsection{Mutual statistical interaction and compact gauge symmetry}
\label{SH}

Consider now an arbitrary loop $C$ on the lattice plane.
Eqs.~(\ref{gauge3}) and (\ref{gauge4}) give \cite{unit}
\begin{eqnarray}
\sum_{\langle i,j \rangle\in C}\, A_{ij}^s &  =  & \pi \sum_{l\in
S_C}\, \left( b^\dagger_{l\uparrow}
b_{l\uparrow}-b^\dagger_{l\downarrow} b_{l\downarrow}\right) \,\, {\rm mod}\, 2\pi\,,\label{plaque1}\\
\sum_{\langle i,j \rangle\in C}\, A_{ij}^h &  =  & \pi \sum_{l\in
S_C}\, h^\dagger_l h_l \,\, {\rm mod}\, 2\pi \,, \label{plaque3}
\end{eqnarray}
where on the left hand side the sum runs over all the links on
$C$\,, while on the right hand side the sum runs over all the sites
inside $C$\,. Eqs.~(\ref{plaque1}) and (\ref{plaque3}) show that the
holon (spinon) carries a $\pi$-flux and couples to the motion of
spinons (holons) via the gauge field $A_{ij}^h$ ($A_{ij}^s$). We see
that the phase string effect effectively ``fractionalizes'' an
electron into two topological objects -- the spinon and the holon --
and the spin and charge degrees of freedom obeys a mutual
statistical interaction. This kind of interaction was also found in
different contexts \cite{Xu09,Galitskii05,DST06}.

It is of conceptual importance that the present theory differs from
the earlier one \cite{KQW05} in the emergence of the local {\it
compact} (instead of non-compact) $U(1)\otimes U(1)$ gauge symmetry,
where the two $U(1)$ gauge degrees of freedom couple to the holon
and the spinon, respectively. Indeed, Eqs.~(\ref{Hj1}), (\ref{Ht1}),
(\ref{plaque1}) and (\ref{plaque3}) are invariant under the
following gauge transformation (All the quantities here may depend
on the imaginary time $\tau$.):
\begin{eqnarray}
\begin{array}{c}
A_{ij}^{s,h} \rightarrow  A_{ij}^{s,h} + \theta_i^{s,h} -
\theta_j^{s,h}\,, \\
b_{i\sigma}^\dagger  \rightarrow  b_{i\sigma}^\dagger \, e^{-i\sigma
\theta_i^h}\,, \qquad b_{i\sigma}  \rightarrow b_{i\sigma} \,
e^{i\sigma \theta_i^h}\,, \qquad h_i^\dagger \rightarrow h_i^\dagger
\, e^{-i \theta_i^s}\,, \qquad b_{i\sigma} \rightarrow b_{i\sigma}
\, e^{i \theta_i^s}\,,
\end{array}
\label{invariance}
\end{eqnarray}
and are apparently invariant upon shifting $A^{s,h}_{ij}$ by
$2\pi$\,. Importantly, due to the single-valued nature the following
mapping
\begin{eqnarray}
\theta^{s,h}: {\cal C}^{s,h} \mapsto U(1) \label{mapping}
\end{eqnarray}
from an arbitrary loop ${\cal C}^{s,h}$ in spacetime lattice to the
gauge group $U(1)$ has the homotopy group $\mathbb{Z}$.


\begin{figure}[h]
  \centering
\includegraphics[width=8cm]{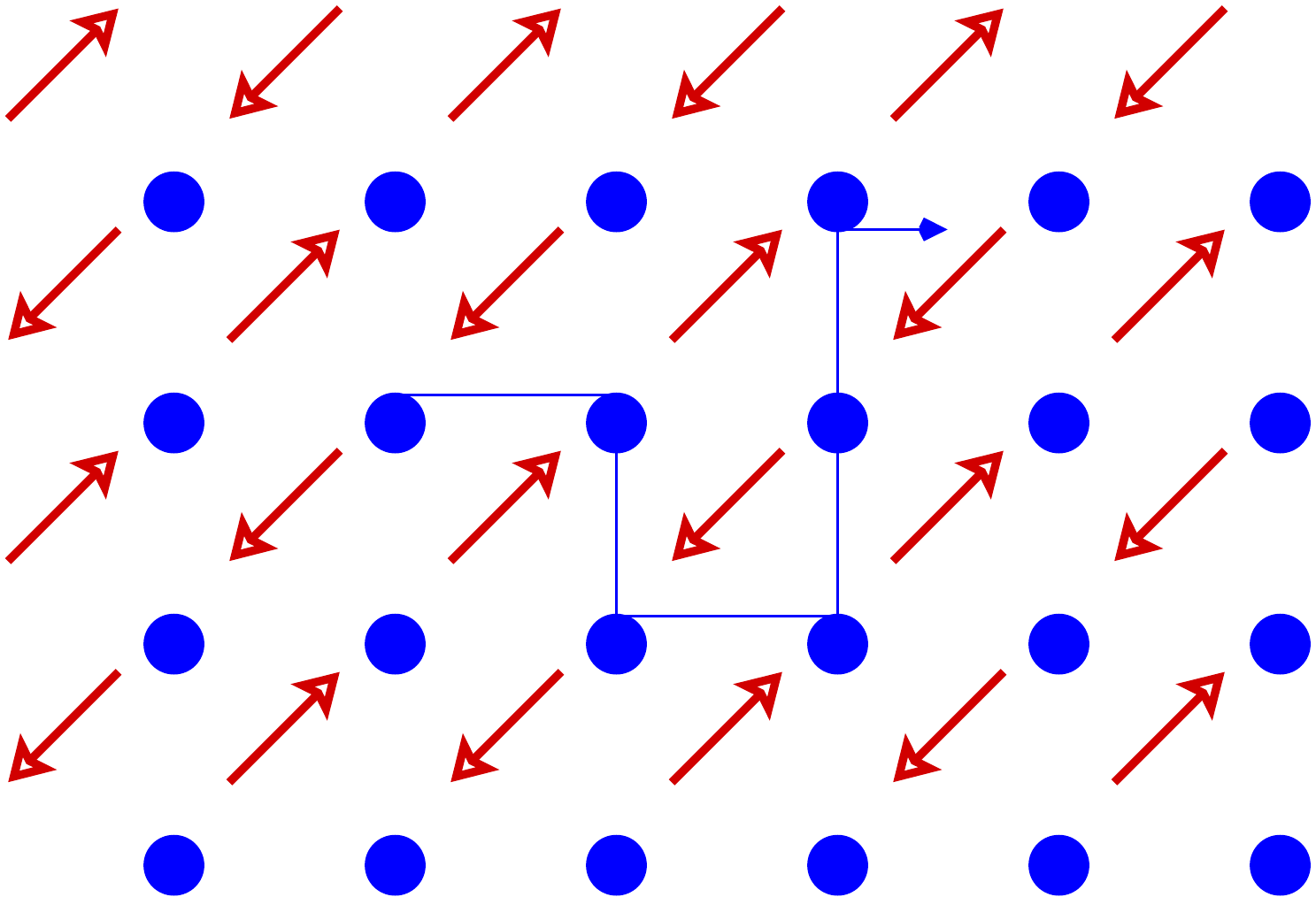}
 \caption{Dual lattice regularization: the lattice (red arrow) where spinons reside is separated from the lattice
 (blue circle) supporting the holon motion (presented by the path).}
  \label{dual}
\end{figure}

\section{Lattice field theory and unconventional order parameters}
\label{sec_latticeMCS}

Our theory has so far been exact. In the remaining of this paper we
will study the mean field approximation of Eqs.~(\ref{Hj1}) and
(\ref{Ht1}), read
\begin{eqnarray}
H_s & = & -J_s \sum_{\langle ij\rangle}\sum_\sigma\, e^{i\sigma \,
A_{ij}^h}\, b_{i\sigma}^\dagger b_{j-\sigma}^\dagger + {\rm h.
c.}\,,
\label{stringH}\\
H_h & = & -t_h \sum_{\langle ij\rangle}\, e^{iA_{ij}^s}\,
h_i^\dagger h_j + {\rm h. c.}\,. \label{Ht2}
\end{eqnarray}
The ``minimal'' Hamiltonian $H_h + H_s $ and Eqs.~(\ref{gauge3}) and
(\ref{gauge4}) constitute the so-called phase string model. It is
important that this model preserves the compact $U(1)\otimes U(1)$
gauge symmetry.

To further proceed we introduce the dual lattice regularization
\cite{KQW05}. Basically, the spinon and the holon reside in their
own square sublattice (Fig.~\ref{dual}). The advantage of this
ultraviolet regularization is as follows: as the constraints
(\ref{plaque1}) and (\ref{plaque3}) are concerned, in the right-hand
side of these two equations, the holon (spinon) does not appear on
the loop $C$ (We stress, however, that this is merely for the
purpose of technical simplicity and violating this condition does
not lead to changes of the results reported in this paper.).
The spinon coordinate system is denoted as $x\equiv (i_1,
i_2,i_0)$, where the first two components refer to the lattice point
in the $x$-$y$ plane and the last to the (discrete) imaginary time.
We set the lattice constant to unity. The holon coordinate system is
denoted as $x\equiv (I_1, I_2,I_0)$, with the origin as
$(-\frac{1}{2}, -\frac{1}{2},0)$ in the holon coordinate system. The
gauge field $A^h_{x+{\hat \mu}x}\equiv A^h_\mu(x)$ is defined on the
spinon lattice, while $A^s_{xx-{\hat \mu}}\equiv A^s_{\mu}(x)$ [as
well as the external electromagnetic field $A^e_{xx-{\hat
\mu}}\equiv A^e_{\mu}(x)$] is defined on the holon lattice. Here, we
denote the unit vector in the $\mu$ direction as $\hat \mu$.
Finally, wherever a lattice function $f(i,I)$ is involved the
relation $(I_1, I_2,I_0)=( i_1+\frac{1}{2},i_2+\frac{1}{2},i_0)$ is
implied for the arguments, and throughout this paper, the Greek
indices $\mu,\nu,\lambda$ run over both the spatial and the temporal
components, while $\alpha,\beta$ only over the spatial component.

Consider the partition function ${\cal Z} \equiv {\rm Tr}
\exp(-\beta {\cal H})$\,. Here, $\beta$ is the inverse temperature
and the Hamiltonian is
\begin{eqnarray}
{\cal H} \equiv -t_h \sum_{I\alpha} \left( e^{i A^s_{\alpha}}
h^\dagger_I h_{I-{\hat \alpha}} + {\rm h.c. } \right) - J_s
\sum_{i\alpha \sigma} \left(e^{i\sigma A^h_{{\alpha}}}
b^\dagger_{i+{\hat \alpha}\sigma} b^\dagger_{i-\sigma} + {\rm h.c. }
\right)
\label{stringH1}
\end{eqnarray}
under the dual lattice regularization. Then, it is a canonical
procedure to divide $\beta$ into ${\cal N}$ pieces, and for each
imaginary time slice we insert the unit resolution. Upon sending
${\cal N}$ to infinity we expect to obtain an expression of ${\cal
Z}$ in terms of the path integral of the configuration of the gauge
fields $A_{\alpha}^{s,h}$ and the matter fields $h,b$ (as well as
their complex conjugates $h^\dagger, b^\dagger$). All the fields are
in turn imaginary time-dependent (which, however, is suppressed to
simplify notations wherever no confusions arise).
We now facilitate this general program by two steps. \\
\\
{\it Step I.} We absorb formally the constraints enforcing on the
gauge fields by the compactness into the measure, obtaining an
expression for the partition function which is
determined by an auxiliary Lagrangian $\tL$\,;\\
\\
{\it Step II.} We work out the constraints explicitly, prompting
$\tL$ to a new Lagrangian $\mL$.

\subsection{Mutual Chern-Simons term}
\label{latticeMCS2}

We begin with Step I. To this end we note that for each imaginary
time slice $\tau=j\beta/{\cal N}\,, j=1,2,\cdots,{\cal N}$\,, the
topological constraints (\ref{plaque1}) and (\ref{plaque3}) imposed
on each plaquette give
\begin{eqnarray}
\begin{array}{c}
\mF_{A^s} \equiv \epsilon^{0\alpha \beta} {d}_\alpha A^s_{\beta}
-\pi \left( b^\dagger_{i\uparrow}
b_{i\uparrow}-b^\dagger_{i\downarrow}
b_{i\downarrow}\right) =0 \,, \\
\mF_{A^h} \equiv \epsilon^{0\alpha \beta} d_\alpha A^h_{\beta} - \pi
h^\dagger_I h_I = 0\,,
\end{array}
\label{plaque2}
\end{eqnarray}
where we define the (spacetime) lattice derivative as $d_\mu f(x)=
f(x+{\hat \mu}) - f(x)$ for the spinon lattice and as $d_\mu f(x) =
f(x) - f(x-{\hat \mu})$ for the holon lattice, with $f(x)$ an
arbitrary lattice function. Taking this into account we find
\begin{eqnarray}
  {\Z} = \int D_{A}[A_{\alpha}^s,A^h_{\alpha}]
D[h^\dagger,h,b^\dagger,b] \delta(\mF_{A^s}) \delta(\mF_{A^h})
e^{-\sum_x {\cal L}_{\rm M}|_{A_0^s=A_0^h=0}}\,, \label{Z}
\end{eqnarray}
with the Lagrangian
\begin{eqnarray}
  {\cal L}_{\rm M} &=& h_I^\dagger (d_0-iA_0^s+\lambda^h)h_I
+t_h \sum_{\alpha} \left( e^{i A^s_{\alpha}} h^\dagger_I h_{I-{\hat
\alpha}} + {\rm c.c. } \right) \nonumber\\
&& + \sum_\sigma b^\dagger_{i\sigma}(d_0-i\sigma
A_0^h+\lambda^s)b_{i\sigma} + J_s \sum_{\alpha \sigma}
\left(e^{i\sigma A^h_{{\alpha}}} b^\dagger_{i+{\hat \alpha}\sigma}
b^\dagger_{i-\sigma} + {\rm c.c. } \right)\,. \label{LM}
\end{eqnarray}
Here ${\rm c.c.}$ stands for the complex conjugate, and the
subscript ``$A$'' in the measure $D_A$ for that as the integral over
the gauge fields are concerned, the constraints enforced by the
compactness (which will be work out explicitly in the next
subsection) are implied.
The parameters $\lambda^{s,h}$ are to be determined below. The
constraint $\delta(\mF_{A^s})\, \delta(\mF_{A^h})$ may be released
by introducing two real number valued fields $A_0^s(x)\,,A_0^h(x)
\in \mathbb{R}$\,. As a result we obtain
\begin{eqnarray}
{\Z} = \int D_A[A_\alpha^s,A_\alpha^h]D_A[A_0^s,A_0^h]
D[h^\dagger,h,b^\dagger,b] \exp\left\{-\sum_x \left({\cal L}_{\rm M}
+ {\cal L}'_{\rm CS} \right)\right\}, \label{Z8}
\end{eqnarray}
where the mutual Chern-Simons term is
\begin{eqnarray}
{\cal L}'_{\rm CS} = \frac{i}{\pi}\left( A_0^s
\epsilon^{0\alpha\beta} d_\alpha A^h_{\beta}+ A_0^h\,
\epsilon^{0\alpha\beta} {d}_\alpha A^s_{\beta}\right) \,.
\label{MCS}
\end{eqnarray}

The exponent in Eq.~(\ref{Z8}) is invariant merely under the {\it
imaginary time-independent transverse} gauge transformation:
\begin{eqnarray}
&& A^h_{\alpha} \rightarrow A^h_{\alpha} + d_\alpha \theta^h
\,,\qquad b_{\sigma} \rightarrow b_{\sigma} e^{i \sigma \theta^h}
\,, \qquad b^\dagger_{\sigma} \rightarrow b^\dagger_{\sigma} e^{-i
\sigma \theta^h}\,,
\nonumber\\
&& A^s_{{\alpha}} \rightarrow A^s_{{\alpha}} + d_\alpha \theta^s
\,,\qquad h \rightarrow h e^{i \theta^s} \,, \qquad h^\dagger
\rightarrow h^\dagger e^{-i\theta^s} \label{transverse2}
\end{eqnarray}
with $\theta^{s,h}$ some regular lattice functions. On the physical
ground we expect that ${\cal L}'_{\rm CS}$ possesses a natural
extension \cite{Altland}:
\begin{eqnarray}
{\cal L}'_{\rm CS} \rightarrow {\cal L}_{\rm CS} = \frac{i}{\pi}
\epsilon^{\mu\nu\lambda}A_\mu^s d_\nu A^h_\lambda \,. \label{MCS1}
\end{eqnarray}
As such, the action: $\sum_x ({\cal L}_{\rm M} + {\cal L}_{\rm CS})$
is invariant under the full local $U(1)\otimes U(1)$ gauge
transformation (\ref{invariance}). In Appendix~\ref{current} we
present an alternative derivation of this Lagrangian.

Note that the elevation: ${\cal L}'_{\rm CS} \rightarrow {\cal
L}_{\rm CS}$ is merely due to the lift of gauge fixing. This is most
easily seen by reversing the procedure. Suppose that we choose the
Coulomb gauge,
the gauge fields are then composed of purely transverse components
which may be explicitly written as \cite{Altland} $A_\perp^s=({d}_y
\theta^s, -{d}_x \theta^s,\phi^s)$ and $A_\perp^h=(d_y \theta^h,
-d_x \theta^h,\phi^h)$ with $\phi^{s,h}$ and $\theta^{s,h}$ some
functions in the spacetime lattice. (The first two components of
$A_\perp^{s,h}$ are the spatial components while the last one the
temporal component.) Inserting this expression into $\sum_x {\cal
L}_{\rm CS}$ we, indeed, end up with $\sum_x {\cal L}'_{\rm CS}$\,.
One must caution that such an elevation may generally result in a
normalization factor of ${\cal Z}$ accounting for the gauge degree
of freedom. However, it does not lead to any physical results and
we, therefore, shall ignore this factor.

To proceed further, we divide the spinon lattice into two
sublattices, say A and B. The spinon field in the sublattice A is
denoted in the same way as before while in the sublattice B is
denoted as $({\bar b}, {\bar b}^\dagger)$. Furthermore, we introduce
a four-component vector, $\Psi_i$ ($i\in {\rm A}$), as follow:
\begin{eqnarray}
\Psi_i \equiv \left(
           \begin{array}{c}
             \psi_i \\
             {\bar \psi}_i \\
           \end{array}
         \right)\,, \qquad \psi_i \equiv \left(\begin{array}{c}
             b_{i\uparrow} \\
             b_{i\downarrow} \\
           \end{array}
         \right)\,, \qquad {\bar \psi}_i \equiv \left(\begin{array}{c}
             \frac{1}{4}({\bar b}^\dagger_{i+\hat x\uparrow} +{\bar b}^\dagger_{i-\hat x\uparrow}
             +{\bar b}^\dagger_{i+\hat y\uparrow}+{\bar b}^\dagger_{i-\hat y\uparrow})\\
             \frac{1}{4}({\bar b}^\dagger_{i+\hat x\downarrow} +{\bar b}^\dagger_{i-\hat x\downarrow}
             +{\bar b}^\dagger_{i+\hat y\downarrow}+{\bar b}^\dagger_{i-\hat y\downarrow}) \\
           \end{array}
         \right)\,.
\label{psi}
\end{eqnarray}
Assuming the spinon fields are smooth over the (spatial) scale of
the lattice constant, the Lagrangian ${\cal L}_M$ is re-expressed as
\begin{eqnarray}
{\cal L}_{\rm M}&=&h^\dagger \left(\hat D_0^s + \lambda^h -t_h
(2+\hat D^s) \right) h + \Psi^\dagger {\cal M}\Psi\,, \label{eq:7}\\
{\cal M}&=& \left(
                    \begin{array}{cccc}
                      \hat D_{0
                      }^h +\lambda^s & 0 & 0 & -J_s (2+(\hat D^h
                      )^*)\\
                      0  & (\hat D_{0
                      }^h)^* +\lambda^s & -J_s (2+\hat D^h
                      ) & 0\\
                      0 & -J_s (2+\hat D^h
                      ) &-(\hat D_{0
                      }^h)^* +\lambda^s& 0\\
                      -J_s (2+(\hat D^h
                      )^*)& 0 &0& -\hat D_{0
                      }^h +\lambda^s
                    \end{array}
                  \right)\,.
\nonumber
\end{eqnarray}
Here, $\hat D_0^s\equiv d_0-iA_0^s$ and $\hat D_{0
}^h \equiv
d_0-i
A_0^h$ are covariant time derivatives.
$\hat D^{s}\equiv \sum_\alpha (d_\alpha-iA_\alpha^{s})^2$ and $\hat
D^{h} \equiv \sum_\alpha (d_\alpha-i
A_\alpha^{h})^2$ are (covariant) discrete Laplacians. (Notice that
all the fields above are non-singular and, therefore, we may perform
a hydrodynamic expansion. Alluding to this, the Laplacians result.)
Finally, the partition function is written as
\begin{eqnarray}
{\Z} = \int D_A[A^s,A^h] D[h^\dagger,h,\Psi^\dagger,\Psi]\,
e^{-\sum_x ({\cal L}_{\rm M}+{\cal L}_{\rm CS})} \,.
\label{Z1}
\end{eqnarray}

\subsection{Lattice field theory}
\label{compact}

\begin{figure}[h]
  \centering
\includegraphics[width=8cm]{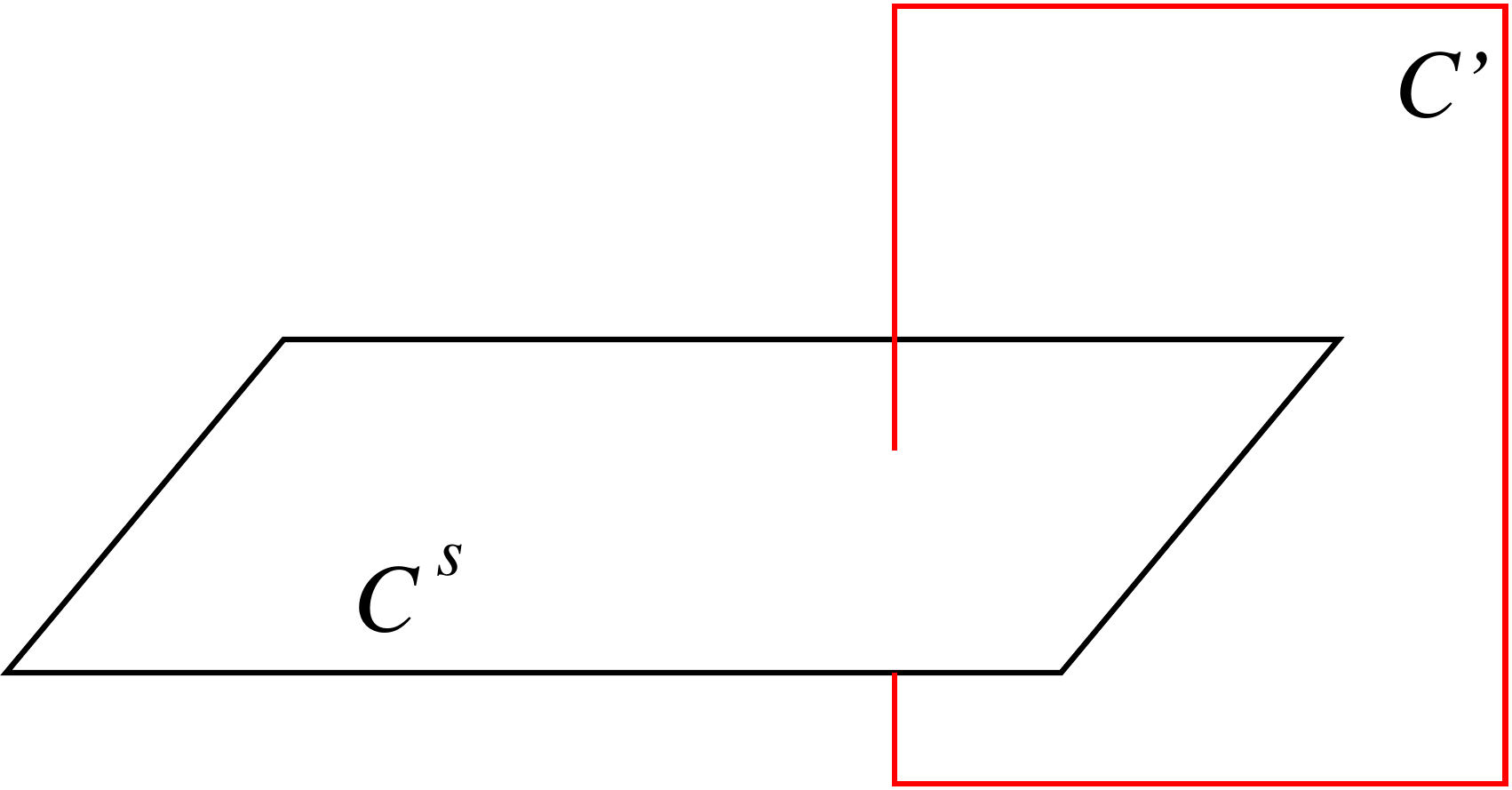}
 \caption{The loop ${\cal C}'$ passes through the area circulated by another loop ${\cal C}^s$.}
  \label{winding}
\end{figure}

So far, the construction of the field theory has been restricted on
the topological trivial sector in the sense that the mapping
(\ref{mapping}) from arbitrarily given loop ${\cal C}^{s,h}$ onto
$U(1)$ has zero winding number. Therefore, the field theory achieved
in Step I does not carry any information regarding the compactness
of the intrinsic gauge group. To fulfill Step II let us start from
analyzing the mapping (\ref{mapping}). It implies that
in the spinon lattice, a loop ${\cal C}'$ may pass through the area
circulated by the loop ${\cal C}^s$ once (Fig.~\ref{winding}) such
that
\begin{eqnarray}
  \epsilon^{\mu\nu\lambda} {d}_\nu {d}_\lambda \theta^{s}(x) =
  2\pi M^s \sum_{x'}J_{{\cal C}'}^\mu(x'), \qquad M^s\in \Bbb{Z} \,.
  \label{compact1}
\end{eqnarray}
Here, the link field $J_{\cal C}^\mu(x)$ takes the value of $ +1$
($-1$) for the link $\langle xx+{\hat \mu}\rangle$ ($\langle x+{\hat
\mu}x \rangle$) on ${\cal C}$ and zero otherwise. Under the gauge
transformation, the action $\sum_x ({\cal L}_{\rm M}+{\cal L}_{\rm
CS})$ is transformed to
\begin{eqnarray}
\sum_x ({\cal L}_{\rm M}+{\cal L}_{\rm CS}) \rightarrow \sum_x
({\cal L}_{\rm M}+{\cal L}_{\rm CS}) + \frac{i}{\pi} \sum_x
\epsilon^{\mu\nu\lambda} {d}_\mu \theta^{s} d_\nu A^h_\lambda =
\sum_x ({\cal L}_{\rm M}+{\cal L}_{\rm CS}) - \frac{i}{\pi} \sum_x
\epsilon^{\lambda \mu\nu} A^h_\lambda {d}_\mu {d}_\nu \theta^{s}\,,
\label{Stransformation}
\end{eqnarray}
where in deriving the second equality we used the integral by parts.
In order for the lattice field theory to be unaffected we demand
that the second term of Eq.~(\ref{Stransformation}) to be multiple
of $2\pi$\,. Taking Eq.~(\ref{compact1}) into account we obtain
\begin{eqnarray}
\sum_{x}\, J_{{\cal C}'}^\mu(x)\, A_{\mu}^h(x) = 0 \, {\rm mod}\,\,
\pi \,. \label{looph}
\end{eqnarray}
Using the Stokes' theorem, we obtain $\sum_x (d_\mu A_{\nu}^h -d_\nu
A_{\mu}^h )\Delta S_x^{\mu\nu}[S_{{\cal C}'}]= 0 \, {\rm mod}\,\,
\pi$ with $\Delta S_x^{\mu\nu}[S_{{\cal C}'}]$ the unit surface
element at $x$\,. (Note that here no summation is implied for
$\mu,\nu$\,.) Since the loop ${\cal C}'$ and thereby the surface
$S_{{\cal C}'}$ are rather general, Eq.~(\ref{looph}) then implies
\begin{eqnarray}
d_\mu A_{\nu}^h - d_\nu A_{\mu}^h = 0 \, {\rm mod}\,\, \pi\,.
\label{fluxspacetimeh}
\end{eqnarray}
That is, the Maxwell tensor of $A^h$ is locally quantized, with a
quanta $\pi$\,. Likewise, we have
\begin{eqnarray}
{d}_\mu A_{\nu}^s - {d}_\nu A_{\mu}^s = 0 \, {\rm mod}\,\, \pi \,.
\label{fluxspacetimes}
\end{eqnarray}

Eqs.~(\ref{fluxspacetimeh}) and (\ref{fluxspacetimes}) are the exact
constraints absorbed in the measure $D_A[A^s,A^h]$ of
Eq.~(\ref{Z1}). To release such constraints we further introduce two
dual integer fields $\mathscr{N}^{h}_{\mu}$ and
$\mathscr{N}^{s}_{\mu}$\,. In doing so we eventually obtain the path
integral representation of the partition function ${\cal Z}$\,,
which is
\begin{eqnarray}
  \mathcal
  {Z}=\sum_{\{\mathscr{N}^{s,h}\}}\int
  D[A^{s},A^{h}]D[h^\dagger,h,\Psi^\dagger,\Psi]\,
  \exp\left\{-\sum_x\left[\mathcal{L}_{\rm M}+\mathcal{L}_{\rm CS}-2i
 \epsilon^{\mu\nu\lambda} \left(A^h_{\mu} d_{\nu} \mathscr{N}^{s}_{\lambda}
 + A^s_{\mu} {d}_{\nu} \mathscr{N}^{h}_{\lambda} \right)\right]\right\}\,.
\label{LGT}
\end{eqnarray}

\subsection{Symmetries}
\label{Sec_SU2}

We turn now to examine various symmetries of the action in
Eq.~(\ref{LGT}). First of all, it is apparently invariant under the
local $U(1)\otimes U(1)$ gauge transformation (\ref{invariance}).
Then, we absorb the integer fields $\mathscr{N}^{s,h}$ into the
mutual Chern-Simons term. In doing so we reduce the Lagrangian in
Eq.~(\ref{LGT}) into
\begin{eqnarray}
\mathcal{L}_{\rm M}+\frac{i}{\pi}\, \epsilon^{\mu\nu\lambda}
\left(A^s_{{\mu}} - 2\pi \mathscr{N}^{s}_{\mu} \right) d_\nu
\left(A^h_{{\lambda}} - 2\pi \mathscr{N}^{h}_{\lambda} \right)\,,
\label{compactaction2}
\end{eqnarray}
where we ignore the irrelevant term multiple of $2\pi$.
From this we see that the field theory is compact in the gauge
degrees of freedom, because a multiple $2\pi$ shift of the gauge
fields is absorbed by the integer fields. Such property is intrinsic
to the discrete lattice geometry but not to the dual lattice
regularization. It is also easy to see that the Lagrangian possesses
both parity and time-reversal symmetry \cite{KQW05}. Under parity
transformation $\mathscr{N}^{h}$ transforms as an axial vector while
$\mathscr{N}^{s}$ as a polar vector. Under time-reversal operation
$\mathscr{N}^{s,h}$ transforms as
\begin{eqnarray}
\mathscr{N}^{s}_0 \rightarrow \mathscr{N}^{s}_0\,,\,
\mathscr{N}^{s}_\alpha \rightarrow - \mathscr{N}^{s}_\alpha \,;
\qquad \mathscr{N}^{h}_0 \rightarrow -\mathscr{N}^{h}_0\,,\,
\mathscr{N}^{h}_\alpha \rightarrow \mathscr{N}^{h}_\alpha \,.
\label{timereversal}
\end{eqnarray}

We further show that the Lagrangian possesses the spin rotation
symmetry. Indeed, transforming $\Psi_i$ as
\begin{eqnarray}
\Psi_i \rightarrow \left(
  \begin{array}{cc}
    e^{i\sigma_3 \Phi_i^h/2}& 0\\
    0&e^{-i\sigma_3 \Phi_i^h/2}\\
  \end{array}
\right) \left(
          \begin{array}{cc}
            U(\theta,\phi,\chi) & 0 \\
            0 & U(-\theta,-\phi,-\chi) \\
          \end{array}
        \right)
\left(
  \begin{array}{cc}
    e^{-i\sigma_3 \Phi_i^h/2}& 0\\
    0&e^{i\sigma_3 \Phi_i^h/2}\\
  \end{array}
\right)\Psi_i\,,
\label{b}
\end{eqnarray}
we find that the Lagrangian is invariant. Note that the $2\times 2$
matrices above are defined on the sector introduced by the doublet
due to the sublattice structure [cf. the definition of $\Psi_i$ in
Eq.~(\ref{psi})], while $U$ and $\sigma_3=\left(
            \begin{array}{cc}
              1 & 0 \\
              0 & -1 \\
            \end{array}
          \right)
$ are defined on the spin sector [cf. the definitions of $\psi_i$\,,
$\bar \psi_i$ in Eq.~(\ref{psi})]. The spin
rotation is
generated by the Euler angles $\theta,\phi$ and $\chi$, i.e.,
\begin{eqnarray}
U(\theta,\phi,\chi) = \left(
                                            \begin{array}{cc}
                                              \cos\frac{\theta}{2}e^{i\phi/2}e^{i\chi/2} & \sin\frac{\theta}{2}e^{-i\phi/2}e^{i\chi/2} \\
                                              -\sin\frac{\theta}{2}e^{i\phi/2}e^{-i\chi/2} & \cos\frac{\theta}{2}e^{-i\phi/2}e^{-i\chi/2} \\
                                            \end{array}
                                          \right) \,.
\label{U}
\end{eqnarray}
Finally, the single-valueness of transformed $\Psi_i$ is guaranteed
by the relation $d_\mu \Phi^h = 2 A^h_\mu$ and
Eq.~(\ref{fluxspacetimeh}). The present theory shows that the spin
rotation symmetry is protected against high-energy
ferromagnetic fluctuations, while the earlier theory \cite{KQW05}
proves the existence of this symmetry only for the low-energy
sector.

To proceed further, we wish to soften the hard-core boson condition,
i.e., Eq.~(\ref{nodouble1}). In doing so, all the above symmetries
must be respected. This goal can be achieved by modifying the
Lagrangian to be
\begin{eqnarray}
\mathscr{L} = \mathcal{L}_{h}+\mathcal{L}_s+\frac{i}{\pi}\,
\epsilon^{\mu\nu\lambda} \left(A^s_{{\mu}} - 2\pi
\mathscr{N}^{s}_{\mu} \right) d_\nu \left(A^h_{{\lambda}} - 2\pi
\mathscr{N}^{h}_{\lambda} \right)\,, \label{compactaction1}
\end{eqnarray}
where the first two terms are the spinon and holon Lagrangian,
respectively, read
\begin{eqnarray}
\begin{array}{c}
{\cal L}_h = h^\dagger \left(\hat D_0^s + \lambda^h -t_h
(2+\hat D^s) \right) h + \frac{u_1}{2} \left(h^\dagger h\right)^2, \\
{\cal L}_s = \Psi^\dagger {\cal M}\Psi + \frac{u_2}{2}
\left(\Psi^\dagger \Psi\right)^2 \,,
\end{array}
\label{Lagrangianh}
\end{eqnarray}
with the last term in ${\cal L}_{h,s}$ describing the on-site
repulsion (the coefficients $u_{1}\gg t_h,\, u_2\gg J_s$), while the
last term in $\mathscr{L}$ [cf. (\ref{compactaction1})] is the
compact mutual Chern-Simons term. Thus, the partition function is
promoted to
\begin{eqnarray}
  \mathcal
  {Z}=\sum_{\{\mathscr{N}^{s,h}\}}\int
  D[A^{s},A^{h}]D[h^\dagger,h,\Psi^\dagger,\Psi]e^{-S}
\label{eq:27}
\end{eqnarray}
with $S=\sum_x \mathscr{L}$\,. Finally, $\lambda^{s,h}$ are
determined by the minimum of the effective action obtained by
integrating out (sum up) all the fields.
Eqs.~(\ref{compactaction1})-(\ref{eq:27}) complete the construction
of our compact mutual Chern-Simons theory. Notice that here
$A_\alpha^{s,h}$ are compact degrees of freedom, i.e.,
$A_\alpha^{s,h}\in [-\pi,+\pi]$, while $A_0^{s,h} \in \Bbb{R}$ not.

It is important to remark that the integer field
$\mathscr{N}^s_\alpha$ ($\mathscr{N}^h_\alpha$) captures the
singular part of the phase fluctuations of the spinon (holon) field.
As we shall see shortly later, in the presence of the spinon (holon)
superfluid $\mathscr{N}^s_\alpha$ ($\mathscr{N}^h_\alpha$) is indeed
the integer field introduced in the well-known Villain's
approximation \cite{Villain}. However, the integer field
$\mathscr{N}^{s,h}_0$ have completely different physical meaning:
(i) Summing up $\mathscr{N}^{s}_0$ leads to the quantization of
$\epsilon^{0\alpha\beta} d_\alpha A_\beta^h$ with a unit $\pi$\,. By
integrating out $A_0^s$\,, we then find $h^\dagger h\in \Bbb{Z}$\,.
Thus, in the limit $u_1\rightarrow \infty$\,, the integral over the
holon field is dominated by the holon configuration where $|h(x)|^2$
takes the value of $0$ or $1$\,. (ii) Similarly, summing up
$\mathscr{N}^{h}_0$ leads to the quantization of
$\epsilon^{0\alpha\beta} d_\alpha A_\beta^s$\,. By integrating out
$A_0^h$\,, we then find $\Psi^\dagger \Sigma \Psi\in
\Bbb{Z}/\{0\}$\,, where $\Sigma\equiv {\rm
diag}(\sigma_3,\,\sigma_3)^T$ is defined in the sector introduced by
the sublattice structure. Thus, in the limit $u_2\rightarrow
\infty$\,, the integral over the spinon field is dominated by the
spinon configuration where for given $x$\,, either
$|b_\uparrow(x)|^2$ or $|b_\downarrow(x)|^2$ takes the value of
unity. Together with the dual lattice regularization, (i) and (ii)
realize the no-double-occupancy constraint (\ref{nodouble}).

\subsection{Unconventional order parameters}
\label{sec_orderparameter}

The Lagrangian (\ref{compactaction1}) shows that the two gauge
fields, $A^{s,h}_\mu$, are dual but the matter fields $\Psi$ and $h$
not. The duality of the gauge fields suggests the introduction of a
pair of Wilson loops, defined as
\begin{eqnarray}
W^{s,h}[{\cal C}]\equiv {\cal Z}^{-1}
\sum_{\{\mathscr{N}^{s,h}\}}\int
D[A^{s},A^{h}]D[h^\dagger,h,\Psi^\dagger,\Psi] e^{-S+i\sum_x
A^{s,h}_\mu J_{\cal C}^{\mu}},
\label{Wilson}
\end{eqnarray}
which depends only on two parameters namely the temperature and the
doping. Here, ${\cal C}$ is a spacetime rectangle with length $T$
($R$) in the imaginary time (spatial) direction.

Physically, the Wilson loop $W^{s}[{\cal C}]$ ($W^{h}[{\cal C}]$)
probes the interaction of a pair of test holons (spinons) at a
distance $R$, $V^h(R)$ ($V^s(R)$), via
\begin{eqnarray}
\begin{array}{c}
  V^h(R)\equiv -\lim_{T\rightarrow \infty} \frac{1}{T}\ln W^{s}[{\cal C}],\\
  V^s(R)\equiv -\lim_{T\rightarrow \infty} \frac{1}{T}\ln W^{h}[{\cal
  C}].
\end{array}
\label{eq:17}
\end{eqnarray}
Furthermore, the analysis in the remaining of this paper suggests
that this pair of Wilson loops suffices to characterize the phase
diagram of the $t$-$J$ model. Therefore, it plays the role of
``order parameter'' and informs the non-Landau-Ginzburg-Wilson
nature of phase transitions involved. We remark that the Wilson
loops introduced here differ crucially from a conventional one
defined on a {\it pure} gauge field theory. In fact, it is the
coupling between the matter and the gauge degrees of freedom that
leads this pair of Wilson loops (potentially) to display very rich
behavior. Physically, the existence of the Wilson loops as order
parameters is the reminiscence of the strategy adopted in quantum
chromodynamics, where the Wilson loop serves as a canonical order
parameter to distinguish the cold and plasma phases
\cite{Giovannangeli01}.

\section{Composition rule for electric transport}
\label{Sec_linearresponse}

In this Section, we will present a microscopic theory of electric
transport basing on the lattice field theory (\ref{compactaction1}).
Specifically, we will derive a so-called composition rule that
expresses the physical electric conductivity tensor,
$\overline{\overline{\sigma}}_e=\{\sigma_e^{\alpha\beta}\}$, in
terms of the holon conductivity tensor,
$\overline{\overline{\sigma}}_h=\{\sigma_e^{\alpha\beta}\}$, and the
spinon conductivity tensor,
$\overline{\overline{\sigma}}_s=\{\sigma_e^{\alpha\beta}\}$. We will
begin with a phenomenological discussion on this rule and then
proceed to a microscopic justification.

\subsection{Phenomenological discussions}
\label{sec_phenomenon}

The minimization of the effective action with respect to gauge
fields leads to the following equations of motion:
\begin{eqnarray}
\begin{array}{c}
\frac{\delta S}{\delta A_\mu^s}\equiv j_{\mu }^{h} +\frac{i}{\pi
}\epsilon ^{\mu \nu \lambda }d_{\nu }\left(
A_{\lambda }^{h}-2\pi \mathscr{N}_{\lambda }^{h}\right)=0, \\
\frac{\delta S}{\delta A_\mu^h}\equiv j_{\mu }^{s} +\frac{i}{\pi
}\epsilon ^{\mu \nu \lambda } d_{\nu }\left( A_{\lambda }^{s}-2\pi
\mathscr{N}_{\lambda }^{s}\right)=0.
\end{array}
\label{eq:1}
\end{eqnarray}
where the spin and charge currents are defined as $j^{s/h}\equiv
\frac{\delta {\cal{L}}_{s/h}}{\delta A^{h/s}}$. From these equations
we obtain
\begin{eqnarray}
j^{h}_\alpha =\frac{1}{\pi}\epsilon^{0\alpha\beta} E^{h}_\beta,
\qquad j^{s}_\alpha =\frac{1}{\pi}\epsilon^{0\alpha\beta}
E^{s}_\beta, \label{eq:3}
\end{eqnarray}
where we have introduced the (macroscopic) internal ``electric''
fields in the imaginary time representation,
\begin{eqnarray}
E_\alpha^{s,h} \equiv -i \epsilon ^{\alpha\mu \nu}d_{\mu }\left(
A_{\nu}^{s,h}-2\pi \mathscr{N}_{\nu}^{s,h}\right). \label{eq:2}
\end{eqnarray}
On the other hand, in the presence of an external electric field
$E^{e}_\alpha$ which couples merely to the holon degree of freedom,
the linear response assumes
\begin{eqnarray}
j^{h}_\alpha(q) =\sigma_h^{\alpha\beta}(q)
[E^{s}_\beta(q)+E^{e}_\beta(q)], \qquad j^{s}_\alpha(q)
=\sigma_s^{\alpha\beta}(q) E^{h}_\beta(q), \label{eq:4}
\end{eqnarray}
where we have passed to the Fourier representation with $q\equiv
(q_x,q_y,i\omega_n)$ the Fourier indices and $\omega_n$ the bosonic
Matsubara frequency. Combining Eqs.~(\ref{eq:3}) and (\ref{eq:4}),
we find that the electric conductivity tensor, defined by
\begin{eqnarray}
j^{h}_{\alpha} =\sigma_e^{\alpha\beta} E^{e}_{\beta}, \label{eq:5}
\end{eqnarray}
obeys the following composition rule:
\begin{eqnarray}
  \overline{\overline{\sigma}}_{e}^{-1}=\overline{\overline{\sigma}}_{h}^{-1}-\pi^{2}
  \epsilon
\overline{\overline{\sigma}}_{s}\epsilon\,,
\label{comb-rule1}
\end{eqnarray}
where $\epsilon=\{\epsilon_{\alpha\beta}\}$ is the antisymmetric
matrix, with $\epsilon_{xy}=-\epsilon_{yx}=1$\,. Note that here and
after, to make formula compact we shall omit the argument $q$
wherever confusions may arise.

\subsection{Microscopic justifications}
\label{sec_RPA}

Now we turn to present a perturbative proof of the composition rule
(\ref{comb-rule1}).
On this purpose we note that in each phase one may in principle
integrate out all the matter fields, arriving at an effective action
of pure gauge fields. Because the holon (spinon) field couples only
to $A^s$ ($A^h$), this gauge field action has the general structure,
\begin{eqnarray}
S_1[A^s+A^e]+S_2[A^h]+\frac{i}{\pi}\sum_x
\epsilon^{\mu\nu\lambda}(A_\mu^s-2\pi \mathscr{N}^s_\mu)d_\nu
(A^h_\lambda-2\pi \mathscr{N}^h_\lambda)\label{eq:10}
\end{eqnarray}
and is gauge invariant, where the first two terms are the kinetic
part. We apply the gauge fixing condition $A^s_0=A^h_0=A^e_0=0$\,
and expand the effective action around its saddle point in terms of
the fluctuating gauge fields $\tilde A^{s,h}_\alpha$. Keeping the
expansion up to the quadratic order, we obtain the fluctuating
action,
\begin{eqnarray}
\frac{1}{2}\sum_{xx'}\left\{[\tilde
A^s_\alpha(x)+A^e_\alpha(x)]\Pi_h^{\alpha\beta}(x-x')[\tilde
A^s_\beta(x')+A^e_\beta(x')] +\tilde
A^h_\alpha(x)\Pi_s^{\alpha\beta}(x-x')\tilde A^h_\beta(x')\right\}
-\frac{i}{\pi}\sum_x \epsilon^{0\alpha\beta}\tilde A_\alpha^s d_0
\tilde A^h_\beta\,, \label{eq:12}
\end{eqnarray}
where $\Pi_{h,s}$ are the polarizations of $\tilde A^s$ and $\tilde
A^h$, respectively.
Integrating out $\tilde A^{s,h}$ eventually reduces the partition
function to ${\cal Z}\propto e^{-S_{\rm eff}[A^e]}$, where the
prefactor is independent of $A^e$ and $S_{\rm eff}[A^e]$ is the
effective action of external gauge fields expanded to the quadratic
order of $A^e$:
\begin{eqnarray}
S_{\rm eff}[A^e]=\frac{1}{2}\sum_{q}A^e_\alpha (q)\Pi_e^{\alpha
\beta}(q)A^e_\beta(q)\,. \label{eq:14}
\end{eqnarray}
Here, $\overline{\overline{\Pi}}_e=\{\Pi_e^{\alpha\beta}\}$ is the
polarization of $A^e$, expressed in terms of
$\overline{\overline{\Pi}}_{s,h}=\{\Pi_{s,h}^{\alpha\beta}\}$ via
\begin{eqnarray}
\overline{\overline{\Pi}}_e^{-1}=\overline{\overline{\Pi}}_h^{-1}-\frac{\pi^2}{\omega_n^2}\epsilon
\overline{\overline{\Pi}}_s \epsilon\,. \label{eq:15}
\end{eqnarray}
Noting $
\sigma_X^{\alpha\beta}=\frac{1}{\omega_n}\Pi_X^{\alpha\beta}
(X=e,h,s)$, we obtained from Eq.~(\ref{eq:15}) the composition rule
(\ref{comb-rule1}).

For later convenience, here we consider a simplification of the
combination rule (\ref{comb-rule1}) by ignoring the off-diagonal
components of the conductivity tensor (namely the crossing
transport). At zero temperature, the static conductivity may be
obtained by taking the limit, $q_\alpha\rightarrow 0$, first and
then $\omega \rightarrow 0$. For an isotropic system the
conductivity tensor is reduced to
$\sigma_X^{\alpha\beta}=\sigma_X\delta_{\alpha\beta}$, and the
composition rule (\ref{comb-rule1}) is reduced to
\begin{eqnarray}
  \sigma_{e}^{-1}=\sigma_{h}^{-1}+\pi
^{2}\sigma_{s}. \label{eq:16}
\end{eqnarray}
We further provide a qualitative explanation of this rule. The
macroscopic electric current (density) $j^h_\alpha$ is fully carried
by holons and driven by both the external electric field $E_\alpha$
and ``electric field'' $E^s_\alpha$ induced by spinons. The latter
finds its origin analogous to that of Ohmic dissipation in type-II
superconductors: each spinon mimics a ``magnetic vortex'' suspending
in holon fluids and, upon moving, generates an electric field {\it
antiparallel} to $j^h_\alpha$, i.e., $E^s_\alpha=-\pi^2 \sigma_s
j^h_\alpha$. In combination with the Ohm's law, i.e.,
$\sigma_h^{-1}j^h_\alpha=-\pi^2 \sigma_s j^h_\alpha+E_\alpha$,
Eq.~(\ref{eq:16}) then follows.

\section{Quantum phase diagram}
\label{sec_phase}

In this Section we shall consider an application of the general
theory developed in Sec.~\ref{sec_latticeMCS} and
\ref{Sec_linearresponse}. Specifically, we will calculate the Wilson
loops, $W^{s,h}[\cal C]$, at zero temperature where such a pair of
non-LGW order parameters depends merely on the doping. We will show
that the order parameter $W^{s}[\cal C]$ ($W^{h}[\cal C]$) displays
non-analyticity where a holon (spinon) deconfinement transition
occurs. This pair of non-LGW order parameters indicates that the
quantum phase diagram (see Fig. 3) is composed of three phases,
namely the antiferromagnetic phase (adjacent to zero doping), the
superconducting phase (far away from zero doping) and a novel phase
-- the Bose-insulating phase (for intermediate doping).

\begin{figure}[h]
  \centering
 \includegraphics[width=13cm]{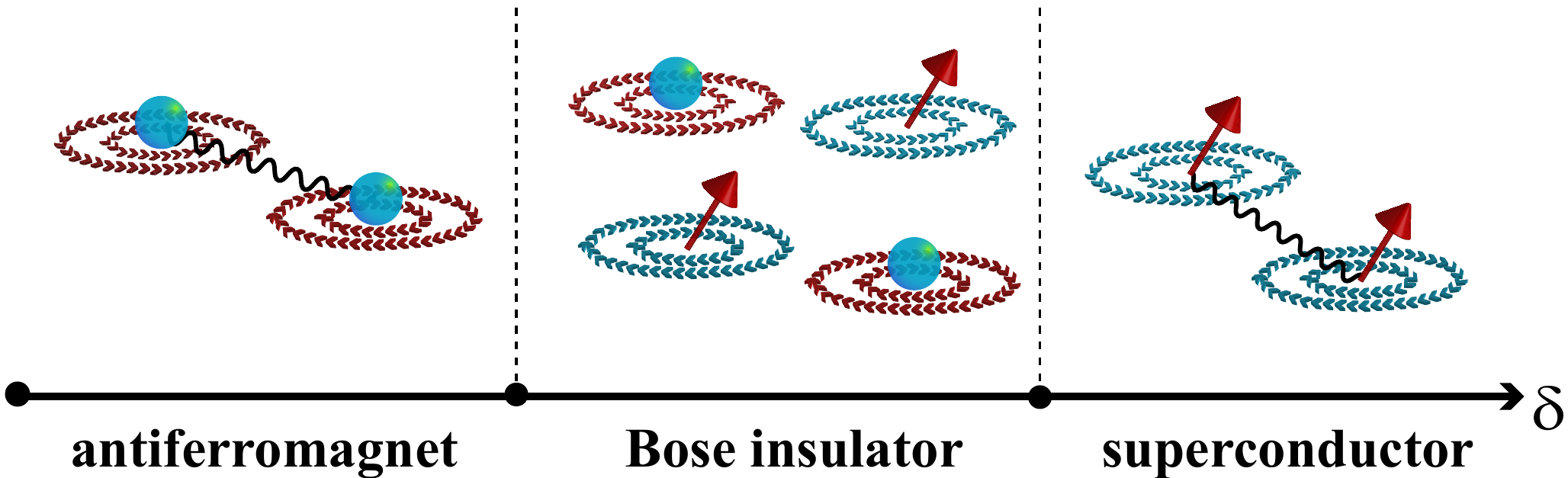}
  \caption{\label{mainresult} (Color online) Quantum phase diagram of underdoped Mott insulators.
  The ball (in blue) and the arrow (in red) stand for the holon and spinon, respectively.
  The vortex, in red (blue), surrounding
  a holon (spinon) arises from the spinon (holon) condensate. The wavy line stands for the confinement.}
\end{figure}

\subsection{
Superconducting phase}
\label{sec_superconducting}

We first study the regime far away from zero doping, where the
superconducting phase is formed. In this phase, the spinons are
confined while holons deconfined. Such peculiar confinement
properties of spinons (holons) are intrinsic to the onset of holon
superfluids. Upon decreasing the doping, the spinons undergo a
deconfinement transition. Correspondingly, the system loses its
superconducting long-ranged order.

\subsubsection{Logarithmic spinon confinement and holon deconfinement}
\label{sec_spin_confinement}

In this regime, the holon Lagrangian ${\cal L}_h$ is minimized upon
the formation of superfluid, justifying the superconducting phase
(with a density $n_h=(2t_h-\lambda^h)/u_1$). In the presence of a
holon superfluid, a uniform ``magnetic field'' is formed and
minimally couples to the motion of spinons [cf.
Eq.~(\ref{Lagrangianh})]. Consequently, a Landau-type gap is opened
in the spinon spectrum and spin excitations are suppressed. For
sufficiently large doping, these spin excitations may be ignored and
the functional integral over the spin degree of freedom $\Psi$ is
frozen at state composing of the so-called resonance valence bond
(RVB) pairs. As such, the functional integral over $\Psi$ accounts
only for an irrelevant overall factor, and the Wilson loop,
$W^{h}[{\cal C}]$\,, is simplified to
\begin{eqnarray}
W^{h}[{\cal C}]\propto \sum_{\{\mathscr{N}^{s,h}\}}\int
D[A^{s},A^{h}]D[h^\dagger,h] \exp\left\{-\sum_x \left[{\cal L}_h+
\frac{i}{\pi} \epsilon^{\mu\nu\lambda} \left(A^s_{{\mu}} - 2\pi
\mathscr{N}^{s}_{\mu} \right) d_\nu \left(A^h_{\lambda} - 2\pi
\mathscr{N}^{h}_{\lambda} \right) -i A^{h}_\mu J_{\cal
C}^{\mu}\right]\right\}.
\label{eq:18}
\end{eqnarray}

To proceed further, we separate the spatial components of the gauge
field, $A^h_{\alpha} - 2\pi \mathscr{N}^{h}_{\alpha}$, into two
parts. The first, denoted as $\bar A^h_{\alpha}$\,, is the
background component which satisfies
\begin{eqnarray}
\epsilon^{0\alpha\beta} d_\alpha \bar A^h_{\beta} = \pi n_h
\label{eq:19}
\end{eqnarray}
and therefore, is imaginary time-dependent. (To be consistent with
this condition it is necessary that $\mathscr{N}^s_0$ must be
uniform in the lattice plane.) The second is the fluctuation
component. Then, we factorize the holon field as
\begin{equation}
h(x)=|h|e^{-i\theta_1(x)}\,,
    \label{eq:42}
\end{equation}
substitute it into Eq.~(\ref{eq:18}) and make the change of
variable: $A_\alpha^s \rightarrow A_\alpha^s-d_\alpha \theta_1$\,.
Integrating out the matter field $h$ with the help of saddle point
approximation and summing up the integer field
$\mathscr{N}^{h}_{\mu}$\,, we obtain
\begin{eqnarray}
W^{h}[{\cal C}] &\propto& \sum_{\{\mathscr{N}^{s}\}}\int
D[A^{s},A^{h}] \exp\left\{-\sum_x
\left[\frac{i}{\pi}\epsilon^{\alpha\nu\lambda} \left(A_\alpha^s-2\pi
\mathscr{N}^{s}_{\alpha}\right)d_\nu A^h_{\lambda} + \frac{i}{\pi}
\epsilon^{0 \alpha\beta} A^s_{0}d_\alpha A_\beta^h \right]\right\}
\exp \left\{i \sum_x A^{h}_\mu
J_{\cal C}^{\mu} \right\} \nonumber\\
&& \qquad \qquad \qquad \qquad \times \exp\left\{-\frac{1}{2u_1}
\sum_x\left(A_0^{s2}+2n_h t_h u_1\,
A_\alpha^{s2}\right)\right\}\,.\label{eq:20}
\end{eqnarray}
Here, the integral variable $A_\alpha^h$ stands for the fluctuation
component of the corresponding gauge field. In deriving this
equation we have used the facts that summing up the fluctuating flux
gives a vanishing result, and that $\mathscr{N}^s_0$ is uniform in
the lattice plane. Notice that an irrelevant factor $e^{i\sum_x \bar
A_\alpha^h J_{\cal C}^\alpha }$ has been omitted. Importantly, the
first term in the first exponent above suggests that in the presence
of holon superfluids, $\mathscr{N}_\alpha^s$ is identical to the
integer fields introduced by the Villain's approximation
\cite{Villain}. Notice that $\theta_1,\, d_\mu \theta_1 \in
[-\pi,+\pi]$ is non-singular, and the singular component of the
phase fluctuations is very captured by (the spatial components of)
the integer field. Integrating out the $A^s$ fields gives
\begin{eqnarray}
W^{h}[{\cal C}] \propto \sum_{\{\mathscr{N}^{s}\}}\int D[A^{h}]
\exp\left\{-\sum_x
\left[\frac{u_1}{2\pi^2}\left(\epsilon^{0\mu\nu}d_\mu
A_\nu^h\right)^2 + \frac{1}{4\pi^2 n_h t_h}\left(\epsilon^{\alpha
\mu\nu}d_\mu A_\nu^h\right)^2 -i A^{h}_\mu \left(J_{\cal
C}^{\mu}+2\epsilon^{\mu\nu\alpha} d_\nu \mathscr{N}^{s}_{\alpha}
\right)\right]\right\}\,.
\label{eq:21}
\end{eqnarray}
From this we see that the dynamics of the fluctuations of the gauge
field $A^h$ displays emergent Lorentz symmetry with a ``speed of
light'' $c_1=\sqrt{2n_h t_h u_1}$\,. By introducing the following
rescaling:
\begin{eqnarray}
(d_0/c_1, d_x, d_y) \rightarrow d_\mu, \qquad (A_0^h/c_1, A^h_x,
A^h_y) \rightarrow A^h_\mu, \qquad (c_1 J_{\cal C}^{0}, J_{\cal
C}^{x}, J_{\cal C}^{y}) \rightarrow J_{\cal C}^{\mu}, \qquad c_1
\mathscr{N}^{s}_\alpha \rightarrow \mathscr{N}^{s}_\alpha,
\label{eq:22}
\end{eqnarray}
we rewrite Eq.~(\ref{eq:21}) as
\begin{eqnarray}
W^{h}[{\cal C}] &\propto& \sum_{\{\mathscr{N}^{s}\}}\int D[A^{h}]
\exp\left\{-\sum_x \left[\frac{1}{4e_1^2}\left(F_{\mu\nu}^h\right)^2
-i A^{h}_\mu \left(J_{\cal C}^{\mu}+2\epsilon^{\mu\nu\alpha} d_\nu
\mathscr{N}^{s}_{\alpha}
\right)\right]\right\}\nonumber\\
&\rightarrow& \sum_{\{\mathscr{N}^{s}\}}\int D[A^{h}]
\exp\left\{-\sum_x \left[\frac{1}{4e_1^2}\left(F_{\mu\nu}^h\right)^2
-i A^{h}_\mu J_{\cal C}^{\mu}-2i A^{h}_0 \epsilon^{0\alpha\beta}
d_\alpha \mathscr{N}^{s}_\beta\right]\right\} \,.\label{eq:23}
\end{eqnarray}
Here, $F_{\mu\nu}^h=d_\mu A^h_\nu-d_\nu A^h_\mu$ is the Maxwell
tensor and $e_1^2=\pi^2/u_1$ is the squared ``bare charge''. In
obtaining the second line we have used the fact, $d_0
\mathscr{N}^{s}_{\alpha}=0$\,, at zero temperature.

If no phase vortices are present, i.e., $2\pi\epsilon^{0\alpha\beta}
d_\alpha \mathscr{N}^{s}_{\beta}=0$\,, Eq.~(\ref{eq:23}) is
simplified to
\begin{eqnarray}
W^{h}[{\cal C}] \propto \int D[A^{h}] \exp\left\{-\sum_x
\left[\frac{1}{4e_1^2}\left(F_{\mu\nu}^h\right)^2 -i A^{h}_\mu
J_{\cal C}^{\mu}\right]\right\}\,.\label{eq:24}
\end{eqnarray}
As we will show in Appendix~\ref{sec_interaction}, it can be further
reduced to (Here, we restore the original unit.)
\begin{eqnarray}
W^{h}[{\cal C}] \propto e^{-TV^s(R)},\qquad V^s(R)=\frac{(e_1
c_1)^2}{2\pi}\ln \frac{R}{R^*_1} =\pi n_h t_h \ln \frac{R}{R^*_1}
\label{eq:28}
\end{eqnarray}
for $T\rightarrow \infty$, where $R^*_1$ is the ultraviolet cutoff.
Notice that the coefficient of the potential, i.e., $\pi n_h t_h$\,,
does not depend on the strength of the on-site repulsive
interaction. Eq.~(\ref{eq:28}) suggests that an external dipole
undergoes logarithmic confinement and has important consequences:
(i) The presence of a pair of free phase vortices of the holon
superfluid, carrying opposite vorticity, is energetically
unfavorable which is consistent with the simplification above; (ii)
In the presence of spin excitations (right, Fig.~\ref{mainresult}),
this dipole may mimic a pair of spinons with opposite or identical
polarizations. In the latter case, a phase vortex with a vorticity
of $\pm 2\pi$ may be excited from the background and bound to a
spinon of flux $\mp \pi$, reversing the sign of the bare {\it
topological} charge of the spinon accordingly, i.e., $\mp \pi
\rightarrow \pm \pi$\,. That is, a pair of spin excitations must be
logarithmically confined.

We turn now to calculate $W^s[{\cal C}]$\,. To this end we may
ignore the fluctuation component of the gauge field $A_\alpha^h$\,.
Following the same procedures of deriving Eq.~(\ref{eq:20}) from
Eq.~(\ref{eq:18}), we find (Here, we do not rescale the gauge field
and $d_\mu$\,.)
\begin{eqnarray}
W^{s}[{\cal C}] \propto \sum_{\{\mathscr{N}^s\}}\int
D[A^{s},A_0^{h}] \exp\left\{-\sum_x
\left[\frac{1}{2u_1}\left(A_0^{s2}+n_h t_h u_1 A_\alpha^{s2}\right)
+ \frac{i}{\pi}\, \epsilon^{0\alpha\beta} \left(A^s_{{\alpha}} -
2\pi \mathscr{N}^{s}_{\alpha} \right) d_\beta A^h_{0} -i A^{s}_\mu
J_{\cal C}^{\mu}\right]\right\}\,, \label{eq:33}
\end{eqnarray}
where we have used the fact that $\theta_1(x)$ is non-singular
giving $d_\alpha d_\beta \theta_1=d_\beta d_\alpha \theta_1$ and
$\sum_x d_\mu\theta_1 J_{\cal C}^\mu=0$\,.
The functional integrals over the temporal and spatial components of
$A^s$ are factorizable, i.e.,
\begin{eqnarray}
W^{s}[{\cal C}] &\propto& \int D[A_0^{s}] \exp\left\{-\sum_x
\left(\frac{1}{2u_1} A_0^{s2} -i A^{s}_0 J_{\cal C}^0
\right)\right\} \nonumber\\
&& \times \int D[A_\alpha^{s}, A_0^h] \sum_{\{\mathscr{N}^s\}}
\exp\left\{-\sum_x \left[\frac{1}{2} n_h t_h A_\alpha^{s2} +
\frac{i}{\pi} \epsilon^{\alpha\beta 0} \left(A^s_{{\alpha}} - 2\pi
\mathscr{N}^{s}_{\alpha} \right) A^h_{0}- i
A^{s}_\alpha J_{\cal C}^{\alpha}\right]\right\} \nonumber\\
&\equiv& \exp\left\{-\left[f_1\left(\frac{c_1T}{\sqrt{n_h
t_h}}\right)+f_2\left(\frac{R}{\sqrt{n_h t_h}}\right)\right]\right\}
\,, \label{eq:30}
\end{eqnarray}
with the scaling function $f_1$ ($f_2$) independent of $R$ ($T$).
This is none but a (holon) deconfinement law -- insensitive to the
details of the scaling functions $f_{1,2}$ -- as expected by holon
superfluids. Shortly we will see that this Wilson loop is not
critical at the spinon deconfinement critical point.

\subsubsection{Non-BCS nature of superconductivity}
\label{sec_conductivity_SC}

Let us subject the system to a small magnetic field with a total
flux $\Phi^e$\,. Consider an area ${\cal A}$ on the spatial plane
that is enclosed by a sufficiently large loop. Since in the ground
state, spinons are confined into RVB pairs and do not contribute a
net flux, integrating out the $A_0^h$ field leads to
\begin{equation}
\sum_{x\in {\cal A}} \Psi^\dagger \Sigma \Psi \equiv \Phi^e \quad
{\rm mod}\, 2\pi\,.
    \label{eq:44}
\end{equation}
As we explained above, the left-hand side is quantized with a quanta
$\pi$\,. This implies in turn that Eq.~(\ref{eq:44}) is none but the
quantization condition of the external magnetic flux, i.e.,
$\Phi^e=n(hc/2e),\, n\in \Bbb{Z}$ in the full unit \cite{unit}.
However, this quantization is intrinsic to the topological spin
excitations and differs in the nature from its counterpart in BCS
superconductivity. In fact, in the absence of the external magnetic
field, i.e., $\Phi^e=0$\,, a single spin carrying a flux of $\pm\pi$
cannot be excited otherwise Eq.~(\ref{eq:44}) is violated. Instead,
spins are excited in pairs, constituting a spin-$0$ (or spin-$1$)
excitation. However, in the presence of the external magnetic field,
i.e., $\Phi^e\neq 0$\,, a single spin can be excited provided it is
nucleated at the magnetic vortex core. Moreover, as a result of the
spin rotation
symmetry, both polarizations are possible \cite{Weng02}.

We then examine the static electric conductivity. As mentioned
above, in the absence of the external magnetic field, spinons are
excited in pairs. If two spinons have identical polarization, they
cannot mobile because they are always bound to a (local) phase
vortex of holon superfluids, as shown in
Sec.~\ref{sec_spin_confinement}. If two spinons have opposite
polarization, as shown in Sec.~\ref{sec_spin_confinement}, they
undergo logarithmic confinement. Therefore, neither of these spinon
pairs supports spinon transport, i.e., $\sigma_s=0$. On the other
hand, since holons undergo Bose condensation, the holon conductivity
is infinite, i.e., $\sigma_h=\infty$. From the composition rule
(\ref{eq:16}) we then obtain $\sigma_e=\infty$. It is important that
according to composition rule, the establishment of
superconductivity relies crucially on a vanishing spinon
conductivity $\sigma_s$ which arises from the spinon confinement. In
other words, it suggests that the disappearance of superconductivity
has to be associated with the spinon deconfinement, where $\sigma_s$
no longer vanishes -- a fact reflecting the non-BCS nature of
superconductivity that we will see in
Sec.~\ref{sec_bose_insulating}.

\subsection{
Antiferromagnetic phase}
\label{sec_antiferromagnetic}

Now we study the regime adjacent to zero doping, where the
antiferromagnetic phase is formed. In this phase, the holons are
confined while spinons deconfined. Such peculiar confinement
properties of spinons (holons) are intrinsic to the onset of spinon
superfluids. Upon increasing the doping, the holons undergo a
deconfinement transition. Correspondingly, the system loses its
antiferromagnetic long-ranged order. The analysis below is largely
parallel to those of Sec.~\ref{sec_superconducting}. Therefore, we
shall only sketch the main steps.

\subsubsection{Logarithmic holon confinement and spinon deconfinement}
\label{sec_holon_confinement}

Consider the case where holons are sufficiently dilute and the holon
Lagrangian is thereby ignored. As a result,
\begin{eqnarray}
W^{s}[{\cal C}] \propto \sum_{\{\mathscr{N}^{s,h}\}}\int
D[A^{s},A^{h}]D[\Psi^\dagger,\Psi] \exp\left\{-\sum_x \left[{\cal
L}_s + \frac{i}{\pi} \epsilon^{\mu\nu\lambda} \left(A^s_{{\mu}} -
2\pi \mathscr{N}^{s}_{\mu} \right) d_\nu \left(A^h_{\lambda} - 2\pi
\mathscr{N}^{h}_{\lambda} \right) -i A^{s}_\mu J_{\cal
C}^{\mu}\right]\right\}\,. \label{eq:34}
\end{eqnarray}
The antiferromagnetism is justified by the existence of homogeneous
saddle points, denoted as $\Psi_0$\,,i.e.,
\begin{eqnarray}
\frac{\delta S}{\delta \Psi^\dagger}\bigg|_{\Psi_0,\,d\Psi_0=0}=0\,.
\label{eq:52}
\end{eqnarray}
In Appendix~\ref{proof} we will show that (i) the saddle point has
the structure as $\Psi_0=(\sqrt{n_\uparrow} e^{i
\theta_\uparrow},\sqrt{n_\downarrow} e^{i
\theta_\downarrow},\sqrt{n_\downarrow} e^{i
\theta_\downarrow},\sqrt{n_\uparrow} e^{i \theta_\uparrow})^T$ with
$n_\sigma$ and $\theta_\sigma$ homogeneous in spacetime and
$n_\uparrow+n_\downarrow=(2J_s-\lambda_s)/(2u_2)$\,;
and (ii) the total spin polarization vanishes, i.e.,
\begin{eqnarray}
\Psi_0^\dagger \Sigma \Psi_0 = 0\,,
    \label{eq:53}
\end{eqnarray}
as a manifestation of antiferromagnetism.

Next we wish to integrate out the spinon field. To this end we
factorize the $\Psi$ field as
\begin{eqnarray}
    \Psi(x)=\left(
              \begin{array}{c}
                \sqrt{n_\uparrow + \frac{\delta n_1(x)+\delta n_2(x)}{2}} \,e^{i
\theta_2(x)/2}\\
                \sqrt{n_\downarrow + \frac{\delta n_3(x)-\delta n_4(x)}{2}} \,e^{-i
\theta_2(x)/2} \\
                \sqrt{n_\downarrow + \frac{\delta n_3(x)+\delta n_4(x)}{2}} \,e^{i
\theta_2(x)/2} \\
                \sqrt{n_\uparrow + \frac{\delta n_1(x)-\delta n_2(x)}{2}}\, e^{-i
\theta_2(x)/2} \\
              \end{array}
            \right)
\label{eq:11}
\end{eqnarray}
and insert it into Eq.~(\ref{eq:34}). The phase field $\theta_2(x)$
generates the Goldstone mode. Furthermore, because of
Eq.~(\ref{eq:53}) the background component of the gauge field
$A_\alpha^s-2\pi \mathscr{N}_\alpha^s$ vanishes. With the help of
the saddle point approximation we obtain from Eq.~(\ref{eq:34})
\begin{eqnarray}
W^{s}[{\cal C}] &\propto& \sum_{\{\mathscr{N}^{h}\}}\int
D[A^{s},A^{h}]D[\delta n,\theta_2] \exp\left\{-\sum_x
\left[\frac{i}{\pi}\epsilon^{\alpha\nu\lambda} \left(A_\alpha^h-2\pi
\mathscr{N}^{h}_{\alpha}\right)d_\nu A^s_{\lambda} + \frac{i}{\pi}
\epsilon^{0 \alpha\beta} A^h_{0}d_\alpha A_\beta^s \right]\right\}
\exp \left\{i \sum_x A^{s}_\mu
J_{\cal C}^{\mu} \right\} \nonumber\\
&& \qquad \qquad \qquad \qquad \times \exp\left\{-\sum_x\left[
-iA_0^h (\delta n_2+\delta n_4)+2J_s (n_\uparrow + n_\downarrow)
(d_\alpha\theta_2-A_\alpha^h)^2\right]\right\}\nonumber\\
&& \qquad \qquad \qquad \qquad \times \exp\left\{\sum_x
\left[\frac{J_s}{4} \left(\delta n_1^2+\delta n_3^2-\delta
n_2^2-\delta n_4^2\right)-\frac{u_2}{2}\left(\delta n_1+\delta
n_3\right)^2\right]\right\}\,.\label{eq:41}
\end{eqnarray}
Again $\theta_2(x)$ is non-singular and instead, the integer field
$\mathscr{N}_\alpha^h$ characterizes the singular part of the phase
field namely the phase vortex of the spinon superfluid. Integrating
out $\delta n$ and $\theta$ gives
\begin{eqnarray}
W^{s}[{\cal C}] \propto \sum_{\{\mathscr{N}^{h}\}}\int
D[A^{h},A^{s}] \exp\left\{-\sum_x \left[\frac{J_s}{8} \left(
A_0^{h2}+16 (n_\uparrow + n_\downarrow) A_\alpha^{h2}\right) +
\frac{i}{\pi}\left( A^h_{{\mu}} - 2\pi \epsilon^{\alpha\nu\lambda}
\mathscr{N}^{h}_{\alpha} \right) d_\nu A^s_{\lambda} -i a^{s}_\mu
J_{\cal C}^{\mu}\right]\right\}\,. \label{eq:35}
\end{eqnarray}
This result is identical to Eq.~(\ref{eq:20}) upon making the
replacement: $u_1^{-1}\rightarrow J_s/4, 2n_h t_h u_1 \rightarrow 16
(n_\uparrow + n_\downarrow)$ for the parameters and $h\rightarrow s,
s\rightarrow h$ for the superscripts. Translating Eq.~(\ref{eq:28})
into the present context, we find
\begin{eqnarray}
W^{s}[{\cal C}] \propto e^{-TV^h(R)},\qquad
V^h(R)=
2\pi (n_\uparrow +
n_\downarrow) J_s \ln \frac{R}{R^*_2} \label{eq:36}
\end{eqnarray}
for $T\rightarrow \infty$\,. Here, $R^*_2$ is the ultraviolet
cutoff. Notice that here the Lorentz symmetry emerges but with a
different ``speed of light''.
The squared ``bare charge'' is also different from that of the
superconducting phase.

Eq.~(\ref{eq:36}) also suggests that an external dipole undergoes
logarithmic confinement and has important consequences as follows.
(i) The presence of a pair of free phase vortices of the spinon
superfluid, carrying opposite vorticity, is energetically
unfavorable. (ii) In the presence of holon excitations, a phase
vortex with a vorticity of $- 2\pi$ may be excited from the
background and bound to a holon of flux $+ \pi$, reversing the sign
of the bare {\it topological} charge of the holon accordingly (the
so-called ``anti-holon''), i.e., $\pi \rightarrow -\pi$.
Eq.~(\ref{eq:36}) then implies that such a holon--anti-holon pair is
bound together via the logarithmic confinement (left,
Fig.~\ref{mainresult}). In other words, two holons are bound to a
phase vortex of the spinon superfluid with a vorticity of $-2\pi$\,.

We turn now to calculate $W^h[{\cal C}]$. Similar to
Eq.~(\ref{eq:33}), we find
\begin{eqnarray}
W^{h}[{\cal C}] \propto \sum_{\{\mathscr{N}^h\}}\int
D[A^{h},A_0^{s}] \exp\left\{-\sum_x
\left[\frac{J_s}{8}\left(A_0^{h2}+16(n_\uparrow+n_\downarrow)
A_\alpha^{h2}\right) + \frac{i}{\pi}\, \epsilon^{0\alpha\beta}
\left(A^h_{{\alpha}} - 2\pi \mathscr{N}^{h}_{\alpha} \right) d_\beta
A^s_{0} -i A^{h}_\mu J_{\cal C}^{\mu}\right]\right\}\,,
\label{eq:38}
\end{eqnarray}
which gives
\begin{eqnarray}
W^{h}[{\cal C}] \propto
\exp\left\{-\left[f_1\left(\frac{c_2T}{\sqrt{n_\uparrow+n_\downarrow}}\right)
+f_2\left(\frac{R}{\sqrt{n_\uparrow+n_\downarrow}}\right)\right]\right\}
\,. \label{eq:39}
\end{eqnarray}
It suggests spinon deconfinement as expected by spinon superfluids,
insensitive to the details of the scaling functions $f_{1,2}$\,. As
we will see below, this Wilson loop in non-critical at the holon
deconfinement critical point.

\subsubsection{Unconventional antiferromagnetism}
\label{sec_conductivity_AF}

Similar to the discussions in Sec.~\ref{sec_conductivity_SC}, the
antiferromagnetism ($\delta\neq 0$) here is unconventional. Let us
subject the system to a spin twist generated by an external gauge
field $A'_\alpha$\,. Since in the ground state, holons and
anti-holons are confined in pairs and do not contribute a net flux,
integrating out the $A_0^s$ field leads to
\begin{equation}
\sum_{x\in {\cal A}} h^\dagger h \equiv \sum_{x\in {\cal
A}}\epsilon^{0\alpha\beta}d_\alpha A'_\beta \quad {\rm mod}\, 2\pi
    \label{eq:45}
\end{equation}
dual to Eq.~(\ref{eq:44}). Since the left-hand side is quantized
(with the same quanta $\pi$), this implies a dual quantization
condition: the external flux generating the spin twist may penetrate
into the antiferromagnet only if it takes the value of $n\pi,\, n\in
\Bbb{Z}$\,. This quantization is intrinsic to the topological holon
excitations. In fact, in the absence of the external spin twist, a
single holon cannot appear in the excitation spectrum otherwise
Eq.~(\ref{eq:45}) is violated. Instead, the holon and the anti-holon
are excited in pairs, constituting a charge-$2$ bosonic excitation.
However, in the presence of the spin twist, a single (anti-)holon
can be excited provided it is nucleated at the center of the spin
twist.

For the static electric transport, we note that in the absence of
external spin twist, the holon and the anti-holon are excited in
pairs. According to Sec.~\ref{sec_holon_confinement}, such pair is
bound to a phase vortex of spinon superfluids. Since the latter is
localized in space, (As such, the spontaneous translational symmetry
breaking appears.) $\sigma_h=0$\,.
From the composition rule (\ref{eq:16}) we then find that the
antiferromagnetic phase is insulating, i.e., $\sigma_e=0$\,. It is
important that such a property of electric transport is intrinsic to
the holon confinement and, therefore, persists to some finite doping
-- the holon deconfinement critical point. This is in sharp contrast
to previous theories \cite{Lee06} where the superconducting phase
was found to be pushed all the way down to zero doping. Finally, it
should be noted that without the integer field
$\mathscr{N}^h_\alpha$ describing the spinon phase vortex, such an
antiferromagnetic insulating phase cannot be established.

\subsection{
Bose insulating phase}
\label{sec_bose_insulating}

The qualitatively different behaviors
of $W^h[{\cal C}]$
described by Eqs.~(\ref{eq:28}) and (\ref{eq:39}) suggest a critical
doping $\delta_2$\,, at which $W^h[{\cal C}]$ -- as a function of
$\delta$ -- is non-analytic. This is the spinon deconfinement
quantum critical point. Likewise, the
expressions of Eqs.~(\ref{eq:30}) and (\ref{eq:36}) for $W^s[{\cal
C}]$
suggest another critical doping $\delta_1$, at which $W^s[{\cal C}]$
is non-analytic. This is the holon deconfinement quantum critical
point. The mechanisms underlying these two quantum critical points
-- the disappearance of the superconducting or the antiferromagnetic
long-ranged order, are independent. Thus, $\delta_1\neq\delta_2$
generally. Since the confinement of holons (spinons) is a
consequence of spinon (holon) condensation, the case of
$\delta_1>\delta_2$ is ruled out. That is, spinons and holons cannot
be confined simultaneously. Instead, we have $\delta_1<\delta_2$
generally. As such, there is an intermediate phase separating the
antiferromagnetic and the superconducting phases.

Indeed, this is a novel phase where both matter fields undergo
condensation characterized by the field $\Psi_0$ ($h_0$). It should
be contrasted to the antiferromagnetic (superconducting) phase where
only the spinon (holon) field $\Psi$ ($h$) is condensed. The fields
$\Psi_0\,,\, h_0$ are determined by the following self-consistent
equations:
\begin{eqnarray}
  && \frac{\delta S}{\delta h^\dagger}\bigg|_{h_0,\,d_0h_0=0}=0\,,
  \quad \frac{\delta S}{\delta \Psi^\dagger}\bigg|_{\Psi_0,\,d_0\Psi_0=0}=0\,,  \label{eq:46}\\
  &&\epsilon^{0\alpha\beta} d_\alpha\left(A_\beta^s-2\pi\mathscr{N}^s_\beta\right)= \pi \Psi^\dagger_0 \Sigma\Psi_0\,, \label{eq:51}\\
  &&\epsilon^{0\alpha\beta} d_\alpha\left(A_\beta^h-2\pi\mathscr{N}^h_\beta\right)=
  \pi h^\dagger_0 h_0\,.
\label{eq:47}
\end{eqnarray}
In general, the field $\Psi_0$ ($h_0$) has an amplitude
inhomogeneous in space. In the Wilson loops (\ref{Wilson}) the
functional integral over $A_\alpha^s\,,\,\mathscr{N}^s_\alpha$
($A_\alpha^h\,,\,\mathscr{N}^h_\alpha$) is dominated by the {\it
small} fluctuations (denoted by $a_\alpha^{s,h}$) near the
background configurations satisfying Eqs.~(\ref{eq:51}) and
(\ref{eq:47}). Taking Eqs.~(\ref{eq:46})-(\ref{eq:47}) into account,
we rewrite Eq.~(\ref{Wilson}) as (To make the formula compact we
denote $A_0^{s,h}$ as $a_0^{s,h}$\,.)
\begin{eqnarray}
W^{s,h}[{\cal C}] &\propto& \int D[a^{s},a^{h}]D[h^\dagger,
h,\Psi^\dagger, \Psi]\exp\left\{-\sum_x \left[{\cal L}_h|_{a_0^s=0}
-ia_0^s (h^\dagger h-h_0^\dagger h_0)\right]\right\}\nonumber\\
&&\times \exp\left\{-\sum_x \left[{\cal L}_s|_{a_0^h=0} -ia_0^h
(\Psi^\dagger \Sigma \Psi-\Psi_0^\dagger \Sigma
\Psi_0)\right]\right\}\exp\left\{-\sum_x
\left[\frac{i}{\pi}\epsilon^{\mu\nu\lambda}a_\mu^s d_\nu a_\lambda^h
- i a^{s,h}_\mu J_{\cal C}^{\mu}\right]\right\}\,.
\label{eq:40}
\end{eqnarray}
The first (second) exponent involves merely $a^s$ ($a^h$). It is
important that integrating out the holon (spinon) field leads to an
effective action of $a^s$ ($a^h$) which is massive. (This can be
readily seen by observing that a spacetime-independent $a^s$ ($a^h$)
cannot be absorbed into the functional integral over $h$ ($\Psi$)
and as such, the effective action of a homogeneous $a^s$ ($a^h$)
must not vanish.) Therefore, the {\it residual} mutual statistical
interaction between gauge field fluctuations, of order ${\cal
O}(d_\mu)$\,, is negligible, and Eq.~(\ref{eq:40}) is simplified to
\begin{eqnarray}
W^{s,h}[{\cal C}]&\propto& \int D[a^{s},a^{h}]D[h^\dagger,
h,\Psi^\dagger, \Psi]\exp\left\{-\sum_x \left[{\cal L}_h|_{a_0^s=0}
-ia_0^s (h^\dagger h-h_0^\dagger
h_0)\right]\right\}\nonumber\\
&& \times \exp\left\{-\sum_x \left[{\cal L}_s|_{a_0^h=0} -ia_0^h
(\Psi^\dagger \Sigma \Psi-\Psi_0^\dagger \Sigma
\Psi_0)\right]\right\}\exp\left[i \sum_x a^{s,h}_\mu J_{\cal
C}^{\mu}\right]\,.
\label{eq:55}
\end{eqnarray}
Integrating out the matter fields then gives
\begin{equation}\label{eq:56}
W^{s,h}[{\cal C}]\propto \int D[a^{s},a^{h}] \exp\bigg\{-\sum_x
\bigg[\frac{1}{2u_1}\left(a_0^{s2}+(c_1 a_\alpha^{s})^2\right) +
\frac{J_s}{8}\left(a_0^{h2}+(c_2 a_\alpha^{h})^2\right) -i
a^{s,h}_\mu J_{\cal C}^{\mu}\bigg]\bigg\}\,.
\end{equation}
From this equation we see that the functional integrals over the
temporal and spatial components of $a^s$ ($a^h$) can be factorized.
As a result, we obtain
\begin{eqnarray}
W^{s}[{\cal C}] &\propto& \exp\left\{-\left[
\tilde f_1\left(\frac{c_1T}{\sqrt{n_ht_h}}\right)+\tilde f_2\left(\frac{R}{\sqrt{n_ht_h}}\right)\right]\right\}\,, \nonumber\\
W^{h}[{\cal C}] &\propto& \exp\left\{-\left[\tilde
f_1\left(\frac{c_2T}{\sqrt{n_\uparrow + n_\downarrow}}\right)+\tilde
f_2\left(\frac{R}{\sqrt{n_\uparrow +
n_\downarrow}}\right)\right]\right\}\,, \label{eq:43}
\end{eqnarray}
where the new scaling functions $\tilde f_{1,2}$ are independent of
$R$ ($T$). Insensitive to the details of these scaling functions,
the Wilson loops (\ref{eq:43}) indicate that an external pair
composed of holon and anti-holon or spinons with opposite
polarizations undergoes deconfinement (middle, Fig.
\ref{mainresult}). Comparing these two expressions with
Eqs.~(\ref{eq:30}) and (\ref{eq:39}), we find that $W^s[{\cal C}]$
($W^h[{\cal C}]$) is indeed non-critical at the spinon (holon)
deconfinement critical point.

The spinon and the holon deconfinement have far reaching
consequences. First, both the long-ranged antiferromagnetic and
superconducting order no longer exist in this phase. Second, recent studies by one of us \cite{Weng11} 
have shown that due to
the spinon and the holon condensation, the phase of the electron
operator is disordered, indicating the existence of a disordered
gapless fermionic mode. As a result, this phase is compressible.
(The present Bose insulator may be considered as a new example of the 
zero-temperature compressible quantum matters proposed recently
\cite{Sachdev11}.) In other words, the holon (charge) density may be
continuously tuned from $\delta_1$ to $\delta_2$\,. Finally,
alluding to $\sigma_{s,h}=\infty$ arising from spinon (holon)
condensation, we find that this phase is also insulating,
$\sigma_e=0$\,, from the composition rule (\ref{eq:16}).

\section{Conclusions}
\label{sec_conclusion}

The $t$-$J$ model (on a bipartite lattice) displays a
non-perturbative sign structure, the so-called phase string effect.
With this effect being taken into full account an electron is
necessarily fractionalized into two bosonic constituents, the holon
(the charge degree of freedom) and the spinon (the spin degree of
freedom). Each constituent is a topological object and carries a
$\pi$ flux. The latter mediates a compact $U(1)$ gauge field,
$A^{s}_\mu$ ($A^{h}_\mu$), minimally couples to the motion of holons
(spinons)--the so-called mutual statistical interaction. In this
work, the exact phase string effect is refined in terms of the
lattice field theory. Based on this field theory, a pair of
unconventional order parameters namely the Wilson loops
$W^{s,h}[{\cal C}]$ is introduced. These two unconventional order
parameters describe the holon (spinon) confinement-deconfinement
property and suffice to characterize the phase diagram of the
$t$-$J$ model in the underdoped regime. We further establish a
general composition rule for the electric transport, which expresses
the electric conductivity in terms of the spinon and the holon
conductivities.

The lattice field theory and the general composition rule are
applied to study the quantum phase diagram. In the antiferromagnetic
regime, where the doping is sufficiently small, spinons undergo
deconfinement while holons are confined, leading to an
(electrically) insulating phase with long range antiferromagnetic
order. Whereas for sufficiently large doping, holons undergo
deconfinement while spinons are confined, leading to a
superconducting phase. (Further analysis of such a superconducting
phase unveils that this is a $d$-wave superconductor and possesses
fermionic Bogoliubov qusiparticles \cite{Weng11}.) We find that
$W^s[{\cal C}]$ ($W^h[{\cal C}]$) displays non-analyticity at some
doping $\delta_1$ ($\delta_2$ with $\delta_1<\delta_2$), where the
system undergoes a holon (spinon) deconfinement quantum phase
transition. Most strikingly, we find that between the
antiferromagnetic and the superconducting phase, there is a novel
phase, the Bose insulating phase. In this phase, despite of spinon
and holon condensation, no long range orders occur and the system is
also electrically insulating. These results inform profoundly
non-Landau-Ginzburg-Wilson nature of quantum phase transitions in
the $t$-$J$ model. We remark that different from the earlier field
theoretic formulation \cite{KQW05}, the present lattice field theory
is compact, and the compactness of the emergent $U(1)$ gauge fields
is essential to the formation of the antiferromagnetic and the Bose
insulating phase. Finally, we should emphasize that the present
theory is not limited to the quantum case. Applications to the
finite temperature case are of fundamental importance and of great
interests, which we leave for future studies.

{\it Note added.} After this paper was submitted, we became aware of
the works by Tesanovic and co-workers \cite{Tesanovic08}. These
authors also found that the duality between the superconducting and
the antiferromagnetic phase dooms to lead to an intermediate
insulating phase. Although this scenario is similar to the quantum
phase diagram presented here, the intermediate phase reported in
these papers has very different nature.

\section*{Acknowledgements}

We thank C. Xu and H. Zhai for very useful discussions. Part of this
work was done during the long-term stay of one of us (C.T.) in
Institut f{\"u}r Theoretische Physik, Universit{\"a}t zu K{\"o}ln.
He is deeply grateful to A. Altland for invaluable support and
encouragement. This work is supported by NSFC grant No. 10834003, by
MOST National Program for Basic Research grant nos. 2009CB929402,
2010CB923003 (P.Y. and Z.Y.W.), by the Alfred P. Sloan foundation
(X.L.Q.), by NSFC grant No. 11174174, by Tsinghua University
Initiative Scientific Research Program, and partly by SFB/TR12 of
the Deutsche Forschungsgemeinschaft (C.T.).

\appendix

\section{Alternative derivation of mutual Chern-Simons term}
\label{current}

In Sec.~\ref{latticeMCS2}, in deriving the mutual Chern-Simons term
the gauge fields $A^{s}_\alpha$ ($A^{h}_\alpha$) are essentially
considered as ``external'' parameters as the holon (spinon) motion
concerned [cf. Eqs.~(\ref{stringH}) and (\ref{Ht2})]. These external
parameters are then subject to the self-consistent constraints
(\ref{plaque2}). In this Appendix, we follow Ref.~\onlinecite{KQW05}
to give an alternative derivation of the mutual Chern-Simons term.

The starting point is the observation that $A^{s,h}_\alpha$
constitute the canonically conjugated pair. Indeed, with the help of
the constraint $\mF_{A^h}=0$ [cf. Eq.~(\ref{plaque2})], the holon
current conservation law
may be rewritten as
\begin{equation}
d_0 \left(\frac{1}{\pi}\, \epsilon^{0\alpha\beta}\, d_\alpha
A^h_\beta\right) + \sum_{\alpha}\, d_\alpha \left(-\frac{\delta
{\cal H}}{\delta A^s_\alpha}\right)=0 \,. \label{contunity1}
\end{equation}
Exchanging the order of the derivatives of the first term, we obtain
\begin{eqnarray}
\frac{1}{\pi} \left(d_x d_0\, A^h_y-d_y d_0\, A^h_x\right) + d_x
\left(-\frac{\delta {\cal H}}{\delta A^s_x }\right) + d_y
\left(-\frac{\delta {\cal H}}{\delta A^s_y} \right)=0 \,,
\label{continuity2}
\end{eqnarray}
which gives
\begin{eqnarray}
\frac{1}{\pi}\, d_0\, A^h_y = \frac{\delta {\cal H}}{\delta A^s_x }
\,, \qquad \frac{1}{\pi}\, d_0\, A^h_x = - \frac{\delta {\cal
H}}{\delta A^s_y} \,. \label{momenta}
\end{eqnarray}
From the spinon current conservation law we may obtain two similar
equations. Together with Eq.~(\ref{momenta}) they justify the
canonically conjugated relation:
\begin{equation}
\Pi^h_x=-\frac{1}{\pi}\, A^s_y\,, \qquad \Pi^h_y=\frac{1}{\pi}\,
A^s_x\,. \label{momenta1}
\end{equation}
Passing to the path integral representation, instead of
Eq.~(\ref{Z8}), we obtain Eq.~(\ref{Z1}) directly.

\section{Derivation of logarithmic confinement}
\label{sec_interaction}

\begin{figure}[h]
  \centering
 \includegraphics[width=8cm]{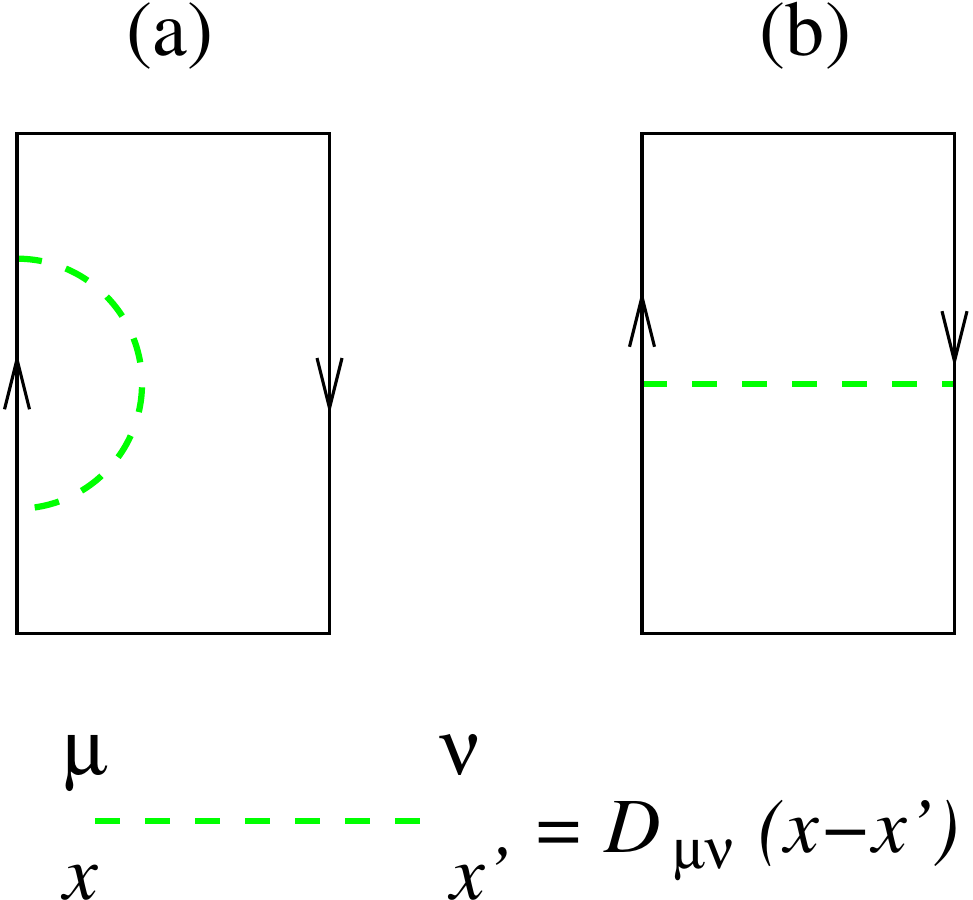}
  \caption{\label{fig_loop} Diagrams contributing to the Wilson loop for $T\rightarrow \infty$.}
\end{figure}

In this Appendix we prove Eq.~(\ref{eq:28}). Throughout this
Appendix we set $c_1=1$\,, and without the loss of generality, we
have set the four corners of the timelike rectangle to be
$(0,0,0),(R,0,0),(R,0,T),(0,0,T)$. Notice that Eq.~(\ref{eq:24}) is
gauge invariant and to further proceed we must fix the gauge. Under
the Feynmann gauge \cite{Peskin}, it becomes
\begin{eqnarray}
W^{h}[{\cal C}] \propto \int D[A^{h}] \exp\left\{-\sum_x
\left[\frac{1}{2e_1^2}\left(d_\mu A_\nu^h\right)^2  -i a^{h}_\mu
J_{\cal C}^{\mu}\right]\right\}\,. \label{eq:31}
\end{eqnarray}
Since we are interested physics at large scales, we pass to the
continuum limit, obtaining the gauge field propagator read
\begin{equation}
D_{\mu\nu}({\bf x}-{\bf x}',\tau-\tau')=\int \frac{d\omega
d^2k}{(2\pi)^3} e^{i{\bf k}\cdot ({\bf x}-{\bf x}')-i\omega
(\tau-\tau')}\frac{e_1^2 \delta_{\mu\nu}}{\omega^2 + |{\bf
k}|^2}=\frac{1}{4\pi}\frac{e_1^2 \delta_{\mu\nu}}{\sqrt {|{\bf
x}-{\bf x}'|^2+|\tau-\tau'|^2}}.\label{eq:32}
\end{equation}
Then, because of $T\gg R$ diagrams with a propagator line starting
from the rectangular sides in the spatial direction are penalized by
(some power of) a small factor $R/T\ll 1$ and thereby are
negligible. As a result, to the leading order expansion in $e_1^2$
Eq.~(\ref{eq:31}) is dominated by the two diagrams shown in
Fig.~\ref{fig_loop}. By ignoring the irrelevant overall numerical
factor, which is the same for both diagrams, they give
\begin{eqnarray}
{\rm diagram \, (a)} &=& -2\times \frac{1}{2} \int D[A^{h}] \int_0^T
d\tau d\tau' A^h_0(0,0,\tau) A^h_0(0,0,\tau')
\exp\left[-\frac{1}{2e_1^2} \sum_x \left(d_\mu A_\nu^h \right)^2
\right]\nonumber\\
&=& - \int_0^T d\tau d\tau' D_{00} (0,0,\tau-\tau') = -
\frac{e_1^2}{4\pi} \int_0^T d\tau d\tau'
\frac{1}{|\tau-\tau'|} \nonumber\\
&=& - \frac{T e_1^2}{2\pi}\int_0^1 dx \ln \frac{xT}{\tau^*},
\label{eq:25}
\end{eqnarray}
with $\tau^*$ an ultraviolet cutoff in the imaginary time direction,
and
\begin{eqnarray}
{\rm diagram \, (b)} &=& 2\times \frac{1}{2} \int D[A^{h}] \int_0^T
d\tau d\tau' A^h_0(0,0,\tau) A^h_0(R,0,\tau')
\exp\left[-\frac{1}{2e_1^2}\sum_x\left(d_\mu A_\nu^h\right)^2
\right]\nonumber\\
&=& \int_0^T d\tau d\tau' D_{00} (R,0,\tau-\tau') = \frac{e_1^2}{4\pi} \int_0^T d\tau d\tau' \frac{1}{\sqrt {R^2+|\tau-\tau'|^2}} \nonumber\\
&=& \frac{e_1^2}{4\pi} \int_0^T d\tau \ln \frac{T-\tau +
\sqrt{(T-\tau)^2+R^2}}{-\tau + \sqrt{\tau^2+R^2}}= \frac{T
e_1^2}{4\pi} \int_0^1 dx \ln \frac{x +
\sqrt{x^2+(R/T)^2}}{-x + \sqrt{x^2+(R/T)^2}}\nonumber\\
&\stackrel{T\rightarrow \infty}{\simeq}& \frac{T e_1^2}{2\pi}
\int_0^1 dx \ln \frac{2Tx}{R}, \label{eq:26}
\end{eqnarray}
respectively. Notice that in the first line of Eqs.~(\ref{eq:25})
and (\ref{eq:26}), the factor $2$ is the combinatorial factor.
Adding these two terms together, we find
\begin{eqnarray}
{\rm diagram \, (a)} + {\rm diagram \, (b)} \stackrel{T\rightarrow
\infty}{\simeq} - \frac{T e_1^2}{2\pi} \ln \frac{R}{2\tau^*}\equiv -
\frac{T e_1^2}{2\pi} \ln \frac{R}{R^*_1}, \label{eq:29}
\end{eqnarray}
which is Eq.~(\ref{eq:28}), with $R^*_1=2\tau^*$ an ultraviolet
cutoff in the spatial direction. Notice that these two diagrams
suffer logarithmical divergences in $T$\,, which, however, cancel
each other exactly upon adding them together.

\section{Spinon superfluids}
\label{proof}

From Eqs.~(\ref{Lagrangianh}) and (\ref{eq:52}) we obtain
\begin{equation}
{\cal M}|_{d_0=A^h=0} \Psi_0 + u_2 (\Psi_0^\dagger
\Psi_0)\Psi_0=0\,.
    \label{eq:54}
\end{equation}
If we re-arrange the components of the vector as follows:
$\Psi_0\equiv (\Psi_{01},\,\Psi_{02},\,\Psi_{03},\,\Psi_{04})^T
\rightarrow (\Psi_{01},\,\Psi_{04},\,\Psi_{02},\,\Psi_{03})^T$\,,
and further introduce two two-component vectors, $z_\sigma,\,
\sigma=\uparrow,\, \downarrow$\,, defined in the sector introduced
by the sublattice structure, then $\Psi_0 = (z_\uparrow^T,\,
z_\downarrow^T)^T$\,. (This two-component structure is defined in
the spin sector.) Consequently, we may rewrite Eq.~(\ref{eq:54}) as
\begin{eqnarray}
  \left(
     \begin{array}{cc}
       \lambda^s & -2J_s \\
       -2J_s & \lambda^s  \\
     \end{array}
   \right) z_\uparrow + u_2 (z_\uparrow^\dagger z_\uparrow + z_\downarrow^\dagger z_\downarrow)z_\uparrow=0\,,
\label{eq:49}\\
  \left(
     \begin{array}{cc}
       \lambda^s & -2J_s \\
       -2J_s & \lambda^s  \\
     \end{array}
   \right) z_\downarrow + u_2 (z_\uparrow^\dagger z_\uparrow + z_\downarrow^\dagger
   z_\downarrow)z_\downarrow=0\,.
\label{eq:48}
\end{eqnarray}
Solving these two equations we
obtain
\begin{eqnarray}
  z_\sigma = \sqrt{n_\sigma} e^{i\theta_\sigma}\left(
                                                             \begin{array}{c}
                                                               1 \\
                                                               1 \\
                                                             \end{array}
                                                           \right)\,,\qquad
                                                           n_\uparrow
                                                           +n_\downarrow=(2J_s-\lambda_s)
                                                           /(2u_2)\,,
  \label{eq:50}
\end{eqnarray}
where $n_\sigma$ and $\theta_\sigma$ are homogeneous in spacetime.
This saddle point solution gives $z_\sigma^\dagger \Sigma_3
z_\sigma=0$ with $\Sigma_3={\rm diag}(1,-1)$ defined on the sector
introduced by the sublattice structure. As a result, we find that
the total spin polarization vanishes namely Eq.~(\ref{eq:53}).


\begin{thebibliography}{}
\bibitem{unit} The spacetime lattice constant, the Planck's constant, the speed of light
in vacuum, and the electron charge are set to unity.
\end{thebibliography}
\end{document}